\documentclass[aps,twocolumn,prd,superscriptaddress,amsmath,amssymb,nofootinbib,longbibliography]{revtex4-2}
\usepackage{graphicx}
\usepackage{dcolumn}
\usepackage{bm}
\usepackage{tikz}
\usetikzlibrary{arrows.meta, calc, decorations.pathmorphing, decorations.markings, patterns} 
\usepackage[dvipsnames]{xcolor}
\definecolor{RoyalPurple}{RGB}{75,0,130}   
\usepackage[colorlinks=true,linkcolor=RoyalPurple,citecolor=RoyalPurple,urlcolor=RoyalPurple]{hyperref}

\usepackage{cancel}
\usepackage[normalem]{ulem}
\usepackage{mathtools}
\usepackage{orcidlink}

\usepackage{staves}
\newcommand{\mystave}{\staveLVII} 
\makeatletter
\let\orig@fnsymbol\@fnsymbol
\def\@fnsymbol#1{\ifnum#1=1 {\color{RoyalPurple}\fontsize{5}{7}\selectfont\mystave}\else\orig@fnsymbol{#1}\fi}
\makeatother

\newcommand{\dGamma}{\delta\Gamma}

\newcommand{\env}{\mathrm{env}}
\AtBeginDocument{%
}

\definecolor{LZbrown}{RGB}{140,90,50}

\begin{document}

\title{A Relativistic Newton's Third Law for Binaries in Environments}

\author{Conor Dyson\,\orcidlink{0000-0002-9742-9422}}
\email{c.dyson.1@bham.ac.uk}

\affiliation{School of Physics and Astronomy \& Institute for Gravitational Wave Astronomy, University of Birmingham, Birmingham, B15 2TT, UK}

\affiliation{Center of Gravity, Niels Bohr Institute, Blegdamsvej 17, 2100 Copenhagen, Denmark}

\author{Adam Pound\, \orcidlink{0000-0001-9446-0638}}
\affiliation{School of Mathematical Sciences and STAG Research Centre,
University of Southampton, Southampton SO17 1BJ, United Kingdom}

\author{Daniel J. D'Orazio\,\orcidlink{0000-0002-1271-6247}}
\affiliation{Space Telescope Science Institute, 3700 San Martin Drive, Baltimore, MD 21218, USA}
\affiliation{Department of Physics and Astronomy, Johns Hopkins University,
3400 North Charles Street, Baltimore, Maryland 21218, USA}
\affiliation{Niels Bohr Institute, Blegdamsvej 17, 2100 Copenhagen, Denmark}

\author{Maarten van de Meent\,\orcidlink{0000-0002-0242-2464}}
\affiliation{Center of Gravity, Niels Bohr Institute, Blegdamsvej 17, 2100 Copenhagen, Denmark}
\affiliation{Max Planck Institute for Gravitational Physics, Am M\"uhlenberg 1, 14476 Potsdam, Germany}

\author{Lorenz Zwick\, \orcidlink{0000-0003-4818-3400}}
\affiliation{Center of Gravity, Niels Bohr Institute, Blegdamsvej 17, 2100 Copenhagen, Denmark}

\date{\today}

\begin{abstract}
We derive the relativistic form of Newton's Third Law governing the interaction between intermediate- and extreme-mass-ratio inspirals and surrounding relativistic matter fields. Working within a two-parameter expansion, we rewrite the evolution of the secondary's orbital elements in a form resembling the integrated forces acting on the environment, avoiding the need to calculate any gravitational perturbations sourced by the dynamical matter content. Using the retarded and advanced Green's functions, we obtain both instantaneous and orbit-averaged laws relating the rate of change of the secondary's orbital energy and angular momentum to that of the matter and the gravitational-wave fluxes at that order. Our results clarify and correct the mixture of ad hoc balance prescriptions currently employed in the literature, providing first-principles forcing functions for the secondary's orbital elements in two ready-to-use forms: a leading-order formulation suitable for fast evaluation with complex relativistic matter simulations, and a first-post-adiabatic formulation of the energy and angular-momentum evolution in environments that are stationary and axisymmetric at leading order.

\end{abstract}

\maketitle

\section{Introduction}
\label{sec:intro}

Extreme-mass-ratio inspirals (EMRIs) are among the flagship sources for the Laser Interferometer Space Antenna (LISA) \cite{2023LRR....26....2A,Colpi:2024xhw}. Composed of a stellar-mass compact object orbiting a massive black hole, they are expected to complete $10^{4}$-$10^5$ strong-field orbits in band, allowing their dynamics to be reconstructed with exceptional precision. This has made them prime targets for tests of strong-field gravity~\cite{LISA:2022kgy}, as well as sensitive probes of dark-matter distributions and scalar fields \cite{Gondolo_1999,Eda:2013gg, Brito:2014wla,Brito:2015oca}. Additionally, the same long-lived inspirals that make them probes of fundamental physics also make them sensitive to perturbations from their astrophysical environment \cite{Barausse:2014tra,Speri:2022}. In particular, a significant fraction of EMRIs are expected to form in active galactic nuclei (AGN), as compact objects are captured into the accretion discs and migrate inward \cite{1991MNRAS.250..505S,Pan:2021ksp}, or as stars and their compact remnants form in situ in the outer regions of the disc before being transported toward the central massive black hole \cite{2003MNRAS.339..937G,2007MNRAS.374..515L,2020ApJ...898...25T,Derdzinski:2022ltb}. In either case, the object enters the LISA band while remaining dynamically coupled to the surrounding gas. Therefore, gas-embedded EMRIs could provide access to a plethora of disc phenomena, including Lindblad resonances \cite{GoldreichTremaine:1979,GoldreichTremaine:1980, Ogilvie02}, stochastic turbulence \cite{2022MNRAS.511.6143Z,2025PhRvD.111j4079C,2026ApJ..1009L...8G} or time-dependent density fluctuations \cite{2025PhRvD.112f3005Z}.

The majority of EMRI modelling work has focused on their vacuum dynamics, with the aim of developing precision waveform models for LISA~\cite{Barack:2018yvs,Pound:2021qin,vandeMeent:2017bcc,Isoyama:2021jjd,Katz:2021yft,Wardell:2021fyy,Nasipak:2023kuf,Chapman-Bird:2025xtd,Nasipak:2025tby,Mathews:2025nyb,Mathews:2025txc,Skoupy:2026ewu, Castillo:2024isq} for tests of general relativity and the black hole paradigm~\cite{LISA:2022kgy,Cardoso:2019rvt}. 

Work has begun to assess the impact of environmental effects on relativistic EMRI dynamics~\cite{Barausse:2007dy,Barausse:2014tra,Speri:2022,Duque:2024mfw}, but realistic or self-consistent models have until recently been lacking.
In recent years, considerable effort has been directed toward building such models, both through fully relativistic local treatments~\cite{Vicente:2025gsg,Vicente:2026dpe} and, in particular, through global calculations of how the matter surrounding a central supermassive black hole responds to the presence of the secondary~\cite{Cardoso:2021wlq,Cardoso:2022whc,Polcar:2022bwv,Duque:2023seg,Brito:2023pyl,Dyson:2025env,Rahman:2025mip,Datta:2025ruh,Li:2025ffh,Polcar:2025yto,Xu:2026aic,Xu:2026cky,Keijzer:2026vul,Li:2026foy,Duque:2025,Hegade:2025a,Dyson:2026ddd,Dittmann:2026rtj}. This effort has produced rapid progress in obtaining consistent solutions for matter in the EMRI's spacetime. However, a rigorous connection between the matter response and the dynamics of the EMRI itself has not yet been established. Existing approaches make this connection by assuming balance laws: generally the secondary's loss of energy and angular momentum are either equated to the fluxes carried to infinity and through the horizon~\cite{Cardoso:2021wlq,Cardoso:2022whc,Duque:2023seg,Brito:2023pyl,Dyson:2025env, Rahman:2025mip, Datta:2025ruh,Li:2025ffh,Xu:2026aic,Xu:2026cky,Keijzer:2026vul}, to the work done on the
matter at specific resonant locations~\cite{Hirata:2011a,Hirata:2011b,Duque:2025,Hegade:2025a,Hegade:2025b}, or to globally integrated matter torques~\cite{Dyson:2026ddd,Dittmann:2026rtj}.

Some of these works also include the additional gravitational-wave (GW) flux terms that arise at second order in the interactions. However, common to all of these approaches is that they compute the work done on, and/or the fluxes of energy and angular momentum carried by, the matter and the gravitational field; they do not rigorously relate these quantities to the work done on the orbiting secondary. Moreover, at second order in the interactions, at which these balance laws are generally applied, further effects enter even for matter fields that are stationary and axisymmetric at leading order: slow-evolution contributions, corrections to the secondary's effective stress-energy from the environment's field, and corrections to the leading-order GW flux due to the environment.

The relevant quantity to calculate the EMRI dynamics is therefore not the work done on the matter by the EMRI, but the work done on the EMRI by the matter including any relativistic GW effects. The direct approach is to compute the gravitational field sourced by the environment and evaluate its (suitably regularised) gradient on the secondary's worldline (see Ref.~\cite{Datta:2026eqj} for a recent self-force calculation of dynamical friction in flat space, and Refs.~\cite{Modrekiladze:2026twz,Wilson-Gerow:2026plu,Bern:2026rdb} for progress with effective-field-theory approaches). Alternatively, one could hope to balance this force against the integrated force applied to the environment, a principle previous works have assumed without derivation. In this paper we make that balance precise, deriving a relativistic analogue of Newton's Third Law for EMRIs evolving self-consistently in matter. When orbit-averaged, the balance law contains the GW flux terms above. Such terms are already present in fluxes computed for EMRIs on exact black-hole--matter spacetimes~\cite{Cardoso:2022whc,Duque:2023seg}, where they appear together with the leading-order vacuum flux, and have since been computed perturbatively for dark-matter spikes~\cite{Rahman:2025mip} and boson clouds~\cite{Keijzer:2026vul,Li:2026foy}, where they dominate the direct matter fluxes by orders of magnitude at small separations. However, their connection to the orbital dynamics has remained ambiguous, and computing them requires gravitational perturbations at mixed EMRI-environment order. The Third Law clarifies how these contributions enter the dynamics and removes the need to calculate any such mixed-order gravitational perturbation. Instead, one takes the stress-energy tensor of a matter solution of choice, from a numerical simulation or a perturbative calculation, and integrates it directly against precomputed point-particle metric perturbations of the Kerr spacetime. Such data are increasingly available through public self-force codes~\cite{Dolan_2022,Dolan:2023enf}, providing a straightforward route to evolving EMRIs embedded in matter.

The structure of the paper is as follows. In Sec.~\ref{sec:newtonian} we present the derivation of the non-relativistic form of Newton's Third Law, which will act as a helpful guide through the following relativistic calculation. In Sec.~\ref{sec:Summary} we summarise the key results of this work. In Sec.~\ref{sec:TwoParamframwork} we provide the two-parameter framework for an EMRI coupled to a general matter stress-energy. In Sec.~\ref{sec:greensandconserv}, guided by the Newtonian calculation, we provide the necessary mathematical structure for our calculation. In particular, we discuss the Green's-function prescription for the linearised Einstein operator and the construction of global hypersurface charges. In Sec.~\ref{sec:BalanceFirst} we derive the relativistic form of Newton's Third Law for asymmetric binaries in spacetimes admitting approximate Killing symmetries. 

The key results of this section, and indeed of the paper as a whole, are the instantaneous balance equation~\eqref{eq:BalanceLaw} and its averaged form Eq.~\eqref{eq:AveragedThirdLaw}. In Sec.~\ref{sec:phasespace} we apply the Third Law, within the fast-to-evolve multiscale formulation~\cite{Pound:2021qin,Mathews:2025nyb}, to what we regard as its two key use cases. The first is the application at leading order to highly dynamical and complex environments in Eq.~\eqref{eq:averagedequationsenvdecomp}, suitable for integration with large-scale numerical simulations. The second is the application at subleading order to more simplified conservative-led environmental setups in Eq.~\eqref{eq:g01_final_1PA}, suitable for perturbative analysis, such as the linear calculations most common in EMRI-environment models under construction for waveform generation at present; this provides a direct means of evolving EMRIs embedded in matter that requires only the relativistic matter response and leading-order metric perturbations of the Kerr spacetime.

Our results can be generically applied to spacetimes admitting approximate Killing symmetries and are therefore easily extended to comparable-mass binary systems in the weak field, which we detail in Appendix~\ref{app:PM_third_law}. For our purposes, however, we are primarily interested in the motion of a compact secondary object in the vicinity of a Kerr black hole, highlighting the symmetries and conserved constants of interest as the orbital energy $\mathcal{E}$ and azimuthal angular momentum $\mathcal{L}_z$.

We use the mostly plus metric signature and geometric units with $G=c=1$. Spacetime components of tensors are denoted by Greek indices.

\section{Motivation and Summary}
\label{sec:Motivation}

A central goal in modelling the interaction between an EMRI and a surrounding matter field is to determine how the matter drives the secondary's orbital evolution. For a Newtonian fluid, this is achieved in the following way: first solving for the fluid's response to the secondary's gravitational field, then solving Poisson's equation for the potential sourced by that response, from which the force on the secondary can be calculated. Newton's Third Law removes the need for the second step. It instead relates the force exerted by the matter on the secondary directly to the force the secondary exerts on the matter, which is more readily available. 

While this connection is easily made in the Newtonian setting, it promises a significant advantage if extended to the relativistic case, where the gravitational potential is replaced by the metric, and Poisson's equation is promoted to the linearised Einstein field equations. A Relativistic Third Law would allow one to forgo the cumbersome calculation of the metric perturbation sourced by the matter field. Hence, we begin by providing the Newtonian derivation of the Third Law, which makes explicit the key properties we will require in the relativistic setting. 

\subsection{Newtonian Case}
\label{sec:newtonian}

We represent the secondary as a point particle with mass $m_{\mathrm{p}}$ moving along a trajectory $\bm{x}_\mathrm{p}(t)$, together with an environment whose mass density is $\rho_{\mathrm{env}}(t,\bm{x})$. Evaluating the force of the matter field on the particle requires computing the gravitational field of the matter, $\Phi_{\mathrm{env}}$, evaluated at the location of the particle,
\begin{equation}
  \bm{F}_\mathrm{p}(t)
  = -m_\mathrm{p}\,\bm{\nabla}\Phi_{\mathrm{env}}\big|_{\bm{x}=\bm{x}_\mathrm{p}(t)}.
  \label{eq:F_p}
\end{equation}
To develop the analogy with the relativistic case, we can write the gravitational field here as
\begin{equation}
  \Phi_{\mathrm{env}}(t,\bm{x})
  = - \int G(\bm{x},\bm{x}')\rho_{\mathrm{env}}(t,\bm{x}')d^3 x',
  \label{eq:Phi_fl}
\end{equation}
where
\begin{equation}
G(\bm{x},\bm{x}')= \frac{1}{\lvert\bm{x}-\bm{x}'\rvert}
  \label{eq:Poisson_Green}
\end{equation}
is the Green's function for the Poisson equation.
The particle applies a force $d\bm{F}_{\mathrm{env}}=-dm_{\mathrm{env}}\,\bm{\nabla}\Phi_\mathrm{p}$ to any given matter element $dm_{\mathrm{env}}=\rho_{\mathrm{env}}d^3x$, in analogy with Eq.~\eqref{eq:F_p}. The integrated force exerted on the entire matter distribution is then
\begin{equation}
  \bm{F}_{\mathrm{env}}(t)
  = -\int \rho_{\mathrm{env}}(t,\bm{x})\,\bm{\nabla}\Phi_\mathrm{p}(t,\bm{x})\,d^3 x.
  \label{eq:F_fl}
\end{equation}
Here the particle's field is given by
\begin{align}
  \Phi_\mathrm{p}(t,\bm{x})
  &= -\int G(\bm{x},\bm{x}')\rho_\mathrm{p}(t,\bm{x}')d^3x' \nonumber\\
  &= -\frac{m_\mathrm{p}}{\lvert \bm{x} - \bm{x}_\mathrm{p}(t)\rvert},
  \label{eq:Phi_p}
\end{align}
where $\rho_\mathrm{p} = m_\mathrm{p}\delta^3(\bm{x}-\bm{x}_\mathrm{p}(t))$ is the particle's mass density.

We now wish to re-express $\bm{F}_\mathrm{p}$, which requires calculating the gravitational field of the environment, in terms of $\bm{F}_{\env}$, which only requires calculating the much simpler particle's field and the matter density profile. 

We do this by first substituting Eq.~\eqref{eq:Phi_fl} into Eq.~\eqref{eq:F_p} to obtain the expression
\begin{align}
  \bm{F}_\mathrm{p}(t) &= m_\mathrm{p}\int \bm{\nabla}_{\bm{x}}G(\bm{x},\bm{x}')\rho_{\mathrm{env}}(t,\bm{x}')d^3x'\bigr|_{\bm{x}=\bm{x}_\mathrm{p}(t)}.
  \label{eq:Fp_step1}
\end{align}
Equivalently, in the more symmetrical form,
\begin{equation}
  \bm{F}_\mathrm{p}(t) = \iint \rho_\mathrm{p}(t,\bm{x})\bm{\nabla}_{\bm{x}}G(\bm{x},\bm{x}')\rho_{\mathrm{env}}(t,\bm{x}')d^3x'd^3x.\label{eq:Fp_step2}
\end{equation}
We now appeal to two key properties of the Green's function. First, 
\begin{equation}\label{eq:Newtonian identity 1}
\bm{\nabla}_{\bm{x}}G(\bm{x},\bm{x}')=-\bm{\nabla}_{\bm{x}'}G(\bm{x},\bm{x}'),
\end{equation}
which follows from $G(\bm{x},\bm{x}')=G(\bm{x}-\bm{x}')$; the Green's function~\eqref{eq:Poisson_Green} only depends on \emph{relative} positions. Second,
\begin{equation}\label{eq:Newtonian identity 2}
G(\bm{x},\bm{x}')=G(\bm{x}',\bm{x}),
\end{equation}
which follows from the fact that Eq.~\eqref{eq:Poisson_Green} only depends on the \emph{distance} between $\bm{x}$ and $\bm{x}'$.
The property~\eqref{eq:Newtonian identity 1} allows us to write Eq.~\eqref{eq:Fp_step2} as
\begin{equation}
    \bm{F}_\mathrm{p}(t) = -\iint \rho_{\mathrm{env}}(t,\bm{x}')\bm{\nabla}_{\bm{x}'}G(\bm{x},\bm{x}')\rho_\mathrm{p}(t,\bm{x})d^3x\,d^3x'.\label{eq:Fp_step3}
\end{equation}
The property~\eqref{eq:Newtonian identity 2} then allows us to write Eq.~\eqref{eq:Fp_step3} as
\begin{align}
  \bm{F}_\mathrm{p}(t)
  =& \int\rho_{\mathrm{env}}(t,\bm{x}')\,
     \bm{\nabla}_{\bm{x}'}\Phi_\mathrm{p}(t,\bm{x}')\,d^3 x'\label{eq:Newton2.9}
  \\
  =&-\bm{F}_{\mathrm{env}}(t).
  \label{eq:Newton3}
\end{align}

Equation~\eqref{eq:Newton3} is precisely Newton's Third Law: the force exerted on the secondary by the environment is equal and opposite to the integrated force exerted on the environment by the secondary. The key feature of this result is that $\bm{F}_\mathrm{p}$ has been expressed entirely in terms of $\Phi_\mathrm{p}$ and $\rho_{\mathrm{env}}$, with no reference to $\Phi_{\mathrm{env}}$. It is this structure that we seek to generalise to the relativistic setting.

This derivation required the key symmetry properties~\eqref{eq:Newtonian identity 1} and \eqref{eq:Newtonian identity 2} of the Green's function for the Euclidean Laplace operator. The fact that those identities do not hold on a curved spacetime implies the breakdown of the intuitive ``equal and opposite reaction'' notion of the Third Law~\cite{Harte:2011ku,Harte:2014wya}. We will find that the curved-spacetime generalisation of those identities gives rise to radiation and time inversion in the relationship between the environment's effect on the particle and the particle's effect on the environment. Moreover, the Newtonian form of the Third Law relies on being able to sum (and compare) forces at different locations, which cannot be straightforwardly done in curved spacetime. Hence, when moving to the relativistic setting, we will replace forces with rates of change of quasi-conserved momenta defined on spatial slices of spacetime. We will then arrive at a surprisingly simple derivation of the Third Law for the charges associated with the Killing symmetries of the background Kerr spacetime.

\subsection{Summary of Results}
\label{sec:Summary}

In the following sections we introduce the formalism and mathematical structures required to derive the relativistic Third Law. Before doing so, we briefly summarise our key findings when applied to the Kerr geometry. Our construction of the Third Law applies to the charges associated with the Killing symmetries of a background spacetime. We therefore focus on the time and azimuthal Killing vectors, $\xi^{\mu}\in\{\partial_t,\partial_{\phi}\}$, with associated energy and angular momentum $P^{(\rm p)}_{\xi} = m_{\rm p}\{-\mathcal{E},\mathcal{L}_z\}$. 

Our first key result, derived in Sec.~\ref{sec:instantaneous}, is the relativistic, instantaneous form of Newton's Third Law~\eqref{eq:Newton3},
\begin{equation}\label{eq:ov_instantaneous}
  \frac{dP^{(\mathrm{p})}_{\xi}}{dt}
  = - \frac{1}{2}\int_{-\infty}^{t} dt'\!\int_{\Sigma_{t'}}
    T^{\alpha'\beta'}_{(\mathrm{env})}\;\mathcal{L}_{\xi'}
       h^{(\mathrm{p,Adv}|t)}_{\alpha'\beta'}\,d\Sigma_0',
\end{equation}
where $h^{(\rm p,Adv|t)}_{\alpha\beta}$ is the advanced metric perturbation generated by the point particle at time $t$, defined in Eq.~\eqref{eq:localised hAdv}. The time-localisation of the particle, together with the explicit integration over its past history, follows from the relativistic extensions of the Green's-function identities~\eqref{eq:Newtonian identity 1}  and \eqref{eq:Newtonian identity 2}, ensuring the appropriate causal connection underlying the Third Law, as discussed in Sec.~\ref{sec:instantaneous}. 

Upon averaging over the orbital time scale, we arrive at the following provocative form of the Third Law at leading order in the environments contribution to the dynamics:
\begin{equation}\label{eq:ov_averaged_flux}
    \left\langle \frac{dP^{(\rm p)}_\xi}{dt}\right\rangle
 = -\left\langle F^{(\rm env)}_\xi\right\rangle
    - \left\langle{\cal F}_\xi^{(\env\times{\rm p})}\right\rangle,
\end{equation}
which, together with the standard point-particle gravitational-wave fluxes, gives the full evolution of the orbital elements at $0$PA. In these expressions, $\langle F^{(\rm env)}_\xi\rangle$ denotes the power (or torque) transferred to the environment by the secondary at leading order, while $\langle \mathcal{F}^{(\env\times\rm p)}_\xi\rangle$ represents a gravitational-wave flux correction arising from the interaction between the matter environment and the secondary, at leading order. These contributions can be expressed explicitly in terms of the \emph{retarded} and \emph{radiative} fields generated by the secondary:
\begin{align}
    \langle F^{(\rm env)}_\xi\rangle
    &= \lim_{T\rightarrow \infty}\frac{1}{4 T}\int_{-T}^{T}dt
    \int_{\Sigma_t} T_{(\env)}^{\mu\nu}
    \mathcal{L}_{\xi} h_{\mu\nu}^{(\rm p,Ret)}\,d\Sigma_0, \\
    \langle \mathcal{F}^{(\env\times\rm p)}_\xi\rangle
    &= -\lim_{T\rightarrow \infty}\frac{1}{2 T}\int_{-T}^{T}dt
    \int_{\Sigma_t} T_{(\env)}^{\mu\nu}
    \mathcal{L}_{\xi} h_{\mu\nu}^{(\rm p,Rad)}\,d\Sigma_0.
\end{align}
The second contribution can, in turn, be expressed explicitly in terms of asymptotic fluxes to infinity and down the primary's horizon, as discussed in Sec.~\ref{sec:averaged_forces}. This provides the necessary correction to a na\"ive extension of the Newtonian Third Law, which would otherwise neglect the additional contributions associated with gravitational radiation. In Sec.~\ref{sec:BalanceTypeD}, we recast these results in phase space, yielding a formulation that is readily applicable to dynamical matter simulations. 

In Sec.~\ref{sec:BalanceTypeC} we then extend the Third Law to first-post-adiabatic ($1$PA) order for environments whose leading-order profiles are stationary and axisymmetric. At $1$PA order we obtain a similar set of forcing functions for the environmental contributions,
\begin{align}\label{eq:ov_g01}
   \left<\frac{d\tilde{P}^{(\rm p)}_\xi}{dt} \right>= -\left<\hat{F}^{(\env)}_{\xi} \right> - \left<\hat{\mathcal{F}}^{(\env\times {\rm p})}_{\xi}\right> -\left<\hat{\mathcal{F}}^{(\mathrm{p}\times \mathrm{p})}_{\xi}\right>,
\end{align}
where the three terms are, respectively, the power (or torque) applied to the environment by the secondary, the GW emission sourced by the perturbed matter, and the correction to the leading-order GW flux from the secondary's motion in the spacetime geometry perturbed by the leading order environment. We present these results in Eqs.~\eqref{eq:1PA_Third_law_kerr_T1}, \eqref{eq:1PA_Third_law_kerr_T2}, \eqref{eq:GW_flux_thrdlaw_1PA_T3}, and \eqref{eq:g01_final_1PA}. Crucially, they are derived from first principles, through the conservation of the secondary's stress-energy tensor and the symmetries of the Green's functions. Evaluating them requires only three ingredients: the matter perturbation induced by the EMRI, $T^{\mu\nu}_{(\env,1,1)}$; the leading-order point-particle metric perturbation, $h^{(\rm p,1,0)}_{\mu\nu}$; and the leading-order metric perturbation of the environment, $h^{(\env,0,1)}_{\mu\nu}$. No gravitational perturbation at mixed EMRI-environment order is needed, which we expect to substantially simplify the calculations required to produce precision EMRI-environment waveform models for LISA.

\section{Formalism and Setup}
\label{sec:Formalism}
Motivated by the Newtonian derivation and having summarised our key results, we next work through the formalism necessary to construct the Third Law in Sec.~\ref{sec:BalanceFirst}. We first present the governing equations of the matter response, subsequently present the relativistic generalisations of Eqs.~\eqref{eq:Newtonian identity 1} and~\eqref{eq:Newtonian identity 2} for Green's functions with Killing symmetries, and finally describe the construction of globally conserved charges. In doing so we will find that the Third Law almost immediately presents itself.

In this section and the remainder of the paper, boldface symbols denote tensors on the fully nonlinear spacetime of the system, rather than Cartesian 3-vectors as they did in the previous section. 
\subsection{Matter Response in the Relativistic Two-Parameter Framework}
\label{sec:TwoParamframwork}
We begin by considering a central Kerr black hole orbited by a compact secondary perturber and an extended matter field environment. In general, incorporating the secondary into the coupled matter-gravity field equations requires a careful procedure of matched asymptotic expansions in the secondary's local neighbourhood, leading to a rigorous prescription for dealing with singularities that arise in the point-particle limit~\cite{Barack:2018yvs}; see Refs.~\cite{Gralla:2008fg,Pound:2012dk,Upton:2021oxf,Musaeus:2026kgp} in the vacuum case, for example. However, to motivate our treatment of the problem, we first imagine we can treat all fields as smooth. We can then consider the field equations in the form
\begin{equation}
  G^{\mu\nu}[{\bf g}] = 8 \pi\left(\epsilon\,T_{(\mathrm{p})}^{\mu\nu}[{\bf g}]
               + \lambda\,T_{(\mathrm{env})}^{\mu\nu}[{\bf g},\{\Psi\}]\right),
  \label{eq:full_ein}
\end{equation}
where ${\bf g}_{\mu\nu}$ is the full spacetime metric and $\{\Psi\}$ denotes the collection of scalar, vector, and/or tensor degrees of freedom describing the environment's
matter content. The parameters $\epsilon$ and $\lambda$ are order-unity counting parameters and track
the strength of the secondary and the environment's gravitational fields respectively. The stress-energy
$T_{(\mathrm{env})}^{\mu\nu}$ describes the matter content of the environment globally, while
$T_{(\mathrm{p})}^{\mu\nu}$ captures the near-field content of the secondary and its surrounding matter. In general, $T_{(\mathrm{p})}^{\mu\nu}$ encompasses not
only the ``bare'' secondary's properties, but also other contributions such as the mass shift due to matter locally bound within the Hill sphere of the secondary. Rates of change of the secondary's multipole moments could, in principle, also account for accretion onto the small body. The matter equations are built from the conservation of the full nonlinear stress-energy tensor, along with other information such as the equation of state. That conservation equation reads 
\begin{equation}\label{eq:conservation}
    {\bm \nabla}_{\mu}\!\left(\epsilon T_{(\rm p)}^{\mu\nu} + \lambda T_{(\env)}^{\mu\nu} \right) = 0,
\end{equation}
 where ${\bm \nabla}_\mu$ is the covariant derivative associated with ${\bf g}_{\mu\nu}$.

In the following sections, we will consider two treatments of the small-$\epsilon$, small-$\lambda$ double limit, and we will derive global balance laws from a local conservation equation of the form~\eqref{eq:conservation}. However, as alluded to above, the small-$\epsilon$ limit introduces singularities in the metric (and therefore in the matter) at the secondary's position, as it effectively shrinks to zero size. The conservation equation~\eqref{eq:conservation}, and the Einstein equation~\eqref{eq:full_ein} from which it is derived, then become ill defined as they involve products of singularities. One can bypass these problems by working consistently away from the particle's worldline, using a puncture scheme~\cite{Pound:2012dk}, in which the secondary is replaced by a singularity in the metric, expressed in terms of the secondary's effective multipole moments derived from the local matching procedure. Alternatively, in some circumstances, one can use classical distribution theory to derive an effective point-particle stress-energy tensor for the secondary~\cite{Gralla:2008fg,Upton:2021oxf,Musaeus:2026kgp}. We discuss appropriate constructions of $T_{(\rm p)}^{\mu\nu}$ in each of the two treatments of the small-$\epsilon$, small-$\lambda$ limit below.

\subsubsection{Nonlinear Matter Response}
\label{sec:nonlinear_response}
At leading order, the matter and the secondary can be treated as sourcing two independent metric perturbations, $h^{(\env)}_{\mu\nu}$ and $h^{(\rm p)}_{\mu\nu}$ respectively. In the two-parameter expansion now common in the EMRI-environments literature~\cite{Rahman:2025mip,Datta:2025ruh,Polcar:2025yto,Brito:2023pyl,Dyson:2025env,Keijzer:2026vul,Li:2025ffh}, one handles nonlinear effects by expanding both the metric and matter fields in powers of $\lambda$ and $\epsilon$. In this section we explore an extension to that approach. Rather than performing a full two-parameter expansion, we expand only the metric degrees of freedom, arriving at a system of nonlinear equations of motion for the environment driven by perturbations in the metric sourced by the secondary. This approach yields a system of equations that do not require the live simulation of the metric but that can still capture strongly nonlinear phenomena in the environment's response. For a thin disc, such effects can include shocks and nonlinear gap formation \cite{Lin:1993}. 

Leaving matter degrees of freedom unexpanded in this way, while still expanding the metric, is the spirit of the ``self-consistent'' approach to self-force theory~\cite{Pound:2009sm,Pound:2012dk,Pound:2015fma,Lewis:2025ydo}. In this approach, we treat the metric as a function of the matter fields and of the secondary's and primary's degrees of freedom, and we then expand it in powers of $\epsilon$ and $\lambda$ while holding all non-gravitational degrees of freedom fixed:  
\begin{equation}\label{eq:self-consistent expansion}
    {\bf g}_{\mu\nu} = g_{\mu\nu} + \sum_{\substack{k,m\geq0\\k+m\geq1}}\epsilon^k\lambda^m h_{\mu\nu}^{(k,m)}(x^\alpha,\gamma,M_A,\{\Psi\}).
\end{equation}
Here $g_{\mu\nu}=g^{\rm Kerr}_{\mu\nu}$, $\gamma$ is the secondary's trajectory, $M_A=(M,S)$ are the primary's mass and spin parameters (which evolve due to absorption through the primary's horizon~\cite{Lewis:2025ydo}), and we suppress dependence on the secondary's multipole moments for simplicity. The coefficients $h_{\mu\nu}^{(k,m)}$ can be obtained by first fixing the spacetime gauge, which puts the Einstein equation in the form 
\begin{equation}\label{eq:relaxed EFE}
G^{\mu\nu}_{\text{(fixed)}}[{\bf g}] =  8\pi \lambda T^{\mu\nu}_{(\env)} \quad \text{away from $\gamma$}, 
\end{equation}
where $G^{\mu\nu}_{\text{(fixed)}}$ is the Einstein tensor with a gauge condition enforced (e.g., the Lorenz gauge or closely related ones~\cite{Pound:2012dk,Pound:2015fma}), and for now we work away from $\gamma$ to avoid the singularities there. Equation~\eqref{eq:relaxed EFE} is said to be ``relaxed'' because $G^{\mu\nu}_{\text{(fixed)}}$ is not subject to the Bianchi identity $\bm{\nabla}_\mu G^{\mu\nu}_{\text{(fixed)}}=0$, meaning Eq.~\eqref{eq:relaxed EFE} can be solved for any matter evolution and any motion of $\gamma$~\cite{Pati:2000vt,Pound:2009sm}. One can hence solve it order by order in $\epsilon$ and $\lambda$ to determine how the metric depends on the matter degrees of freedom, without simultaneously solving for the matter. 

The matter evolution is then determined from conservation of stress-energy (together with any other matter equations), and imposing conservation enforces that the solution ${\bf g}_{\mu\nu}$ to the relaxed Einstein equation also satisfies the non-relaxed one to the same order~\cite{Pound:2012dk}. First consider the conservation equation~\eqref{eq:conservation} away from the secondary, where
\begin{equation}\label{eq:divTenv=0}
    {\bm \nabla}_{\mu}T^{\mu\nu}_{(\env)}[{\bf g}, \{\Psi\}] = 0  \quad \text{away from $\gamma$}.
\end{equation}
Having expanded the metric to solve the relaxed Einstein equation, we do not expand the matter fields $\{\Psi\}$ when solving Eq.~\eqref{eq:divTenv=0}. Instead, we substitute the metric expansion~\eqref{eq:self-consistent expansion}, at whatever are the highest known orders of $\epsilon$ and $\lambda$, into Eq.~\eqref{eq:divTenv=0} to obtain a nonlinear equation for the unexpanded fields $\{\Psi\}$. This is the same strategy used to derive self-consistent equations of motion for the secondary's trajectory $\gamma$. In this way, we allow $\{\Psi\}$ to respond fully nonlinearly to ${\bf g}_{\mu\nu}$.

To explain this scheme more concretely, we consider our main case of interest: the leading interaction between the secondary's metric perturbation and the matter. We truncate Eq.~\eqref{eq:self-consistent expansion} at linear order, giving us the spacetime as linearly perturbed by the secondary and matter,
\begin{equation}\label{eq:g linear}
    {\bf g}_{\mu\nu} = g_{\mu\nu} + \epsilon\, h^{(\rm p)}_{\mu\nu} + \lambda\, h^{(\rm env)}_{\mu\nu},
\end{equation}
with $h^{(\rm p)}_{\mu\nu}= h^{(1,0)}_{\mu\nu}(x^\alpha,\gamma)$ and $h^{(\rm env)}_{\mu\nu} = h^{(0,1)}_{\mu\nu}(x^\alpha,\{\Psi\})$, suppressing dependence on the primary's mass and spin for simplicity. The particle's trajectory $\gamma$ is unspecified here, such that it can freely respond to the environment. Substituting the metric~\eqref{eq:g linear} into the conservation equation~\eqref{eq:divTenv=0} and truncating at order $\epsilon\lambda$, neglecting order-$\lambda^2$ self-gravity effects, 
we obtain a matter equation given by
\begin{align}\label{eq:Nonlinearmatter}
    \nabla_{\mu}T^{\mu\nu}_{(\env)}[g, \{\Psi\}] = &-\dGamma^{\mu}_{\mu\gamma}[h^{(\rm p)}]\,T^{\gamma\nu}_{(\env)}[g,\{\Psi\}]
   \notag \\& -\dGamma^{\nu}_{\mu\gamma}[h^{(\rm p)}]\,T^{\mu\gamma}_{(\env)}[g,\{\Psi\}]
   \notag \\&- \nabla_{\mu}\,\delta^{(1,0)}T^{\mu\nu}_{(\env)}[h^{(\rm p)}, \{\Psi\}]
\end{align}
away from $\gamma$, where covariant derivatives without further decoration are those of the background $g_{\mu\nu}$, and we have defined
\begin{align}
&\delta^{(k,m)} T^{\mu\nu}_{(\env)}[h,\{\delta\Psi\}]:= \\ \notag
&\frac{1}{k!\,m!}\frac{d^{k}}{ds^{k}}\frac{d^{m}}{du^{m}}\big\{T^{\mu\nu}_{(\env)}[g + s h,\{\Psi\}+ u\{\delta \Psi\}]\big\}_{s=u=0},
\end{align}
and
\begin{equation}
\dGamma^{\gamma}_{\mu\nu}[h] := \frac{1}{2}g^{\gamma\beta}\left( \nabla_{\mu}h_{\beta\nu}+\nabla_{\nu}h_{\beta\mu} - \nabla_{\beta}h_{\mu\nu}\right).
\end{equation}

Equation~\eqref{eq:Nonlinearmatter} is the starting point for calculating the nonlinear response of matter in curved space perturbed by the secondary's field. With input data for $h^{(\mathrm{p})}_{\mu\nu}$ from self-force theory, it describes the matter dynamics in the presence of a fully relativistic EMRI while circumventing the need to simulate the metric evolution simultaneously with the matter. If one is only interested in this matter response, one need not compute the linear metric perturbation $h^{(\rm env)}_{\mu\nu}$ sourced by the (nonlinear) matter fields, but we will require this linear perturbation in later sections in order to deduce the backreaction of the matter on the particle's trajectory.

Importantly, the construction presented here relies on two key assumptions. First, the metric only responds to the matter in a weakly nonlinear, perturbative way; a source of order $\epsilon^k\lambda^m$ generates a metric perturbation of order $\epsilon^k\lambda^m$, justifying Eq.~\eqref{eq:self-consistent expansion}. Second, the matter's strongly nonlinear response to the metric is still well ordered, in the sense that higher-order metric perturbations, and higher powers of metric perturbations, provoke smaller responses. We expect these assumptions to hold generally for the mass ratios and separations relevant to EMRIs in the LISA band. 

Even given those assumptions, subtleties arise from the singularity at the particle. Equations~\eqref{eq:relaxed EFE}, \eqref{eq:divTenv=0}, and~\eqref{eq:Nonlinearmatter} only hold at points away from the particle, while in later sections we require integrals over regions that include $\gamma$ as well as local stress-energy conservation at $\gamma$. The singularity in $h^{(\rm p)}_{\mu\nu}$ could spawn a highly singular matter field through the nonlinear equation~\eqref{eq:Nonlinearmatter}, with rapid variations on small scales that potentially prevent one from applying the method of matched asymptotic expansions (which traditionally assumes there is no evolution on the time scale of the secondary or on spatial scales smaller than the secondary). Deriving a puncture or an effective point-particle stress-energy tensor could then be challenging. Moreover, our integrals in later sections will diverge for a divergent solution to Eq.~\eqref{eq:Nonlinearmatter}. A proper treatment of this is an important direction for future work, but one we do not consider here, with potential implications for physically motivated sink prescriptions in hydrodynamical simulations~\cite{Duffell:2019uuk}.

In this section we assume that singularities are avoided using a softened particle whose field returns the full
retarded field outside a specified softening radius $\beta$ and is finite inside it. Softening divergent potentials is
commonplace in nonlinear simulations of binary systems in matter~\cite{Price2007,Iannuzzi:2011,Rowan:2025ppu}; however,
the approaches applied in the Newtonian context rely on the point-mass potential being harmonic away from the
particle, a property the metric perturbation does not have in curved spacetime.

In curved space we can instead soften only the local
expansion of the singular part of the secondary's field, commonly referred to in the self-force literature as the
puncture field, given by
\begin{equation}\label{eq:puncture_expansion}
    h^{\mathcal P}_{\mu\nu}\sim \frac{2m_{\rm p}}{r}\delta_{\mu\nu}+ \cdots
\end{equation}
in the particle's rest frame, where $r$ in this context is the local proper distance to the particle. Owing to the structure of the first-order puncture~\cite{Pound:2012dk}, it can be shown to be polyharmonic,\footnote{ Each term in the first-order puncture satisfies $(\nabla^{2})^{n}h^{\mathcal P}_{\mu\nu}=0$ away from the particle for some integer $n$, where $\nabla^{2}$ is the flat-space Laplacian in local coordinates centred on the particle. \label{fn:polyharmonic}} so it is possible to define the softened field by mollifying only the puncture component,
\begin{equation}\label{eq:soft}
    h^{(\rm soft\,p)}_{\mu\nu}=h^{\rm Res}_{\mu\nu}+\int h^{\mathcal P}_{\mu\nu}(\bm x')\,\psi_\beta(|\bm x-\bm x'|)\,d^3x',
\end{equation}
where $h^{\rm Res}_{\mu\nu}:=h^{(\rm p)}_{\mu\nu}-h^{\mathcal P}_{\mu\nu}$ is the residual field, and $\psi_{\beta}$ is a mollifier which is bounded, normalised to unity, and vanishes outside a softening radius $\beta$. We expect this form of softening to lead to a well-defined setup, similar to those used in the Newtonian setting, in which the softened field returns precisely to the original field outside $\beta$. We leave the practical details, including the requirements on the moments of the mollifier, to future work focused on implementation.

If we extend this softening through to higher orders in $\epsilon$ and $\lambda$, we can then extend all the fields to $\gamma$ and define the particle's softened stress-energy tensor as
\begin{equation}\label{eq:Tsoft nonlinear}
    \epsilon T^{\mu\nu}_{(\text{soft p})} := \frac{1}{8\pi} G^{\mu\nu}_{(\rm fixed)}\bigl[{\bf g}^{(\text{soft})}\bigr] - \lambda T^{\mu\nu}_{(\env)}, 
\end{equation}
where ${\sf g}^{(\text{soft})}_{\mu\nu}=g_{\mu\nu} + \epsilon h^{(\text{soft p})}_{\mu\nu}+\lambda h^{(\env)}_{\mu\nu}+\cdots$ is the softened version of the full metric~\eqref{eq:self-consistent expansion}. 
${\sf g}^{(\text{soft})}_{\mu\nu}$ then satisfies the full, non-relaxed Einstein equation if and only if the stress-energies satisfy the conservation equation
\begin{equation}\label{eq:Tsoft conservation}
    \bm{\nabla}^{(\text{soft})}_\mu \left(\epsilon T^{\mu\nu}_{\text{(soft p)}}+\lambda T^{\mu\nu}_{(\env)}\right) = 0.
\end{equation}
At leading order, Eq.~\eqref{eq:Tsoft nonlinear} reduces to
\begin{equation}\label{eq:Tp soft linear}
T^{\mu\nu}_{(\text{soft p})} = \frac{1}{8\pi}\delta G^{\mu\nu}_{(\rm fixed)}[h^{(\text{soft p})}],
\end{equation}
where $\delta G^{\mu\nu}_{(\rm fixed)}$ is the (gauge-fixed) linearised Einstein tensor on the Kerr background. This leading term vanishes outside of the softening radius and limits to the standard stress-energy tensor in the limit $\beta\to0$. Equation~\eqref{eq:Tsoft conservation} then reduces back to Eq.~\eqref{eq:Nonlinearmatter} at order $\epsilon\lambda$ outside the softening radius.

This softening procedure ensures well-defined equations and local stress-energy conservation. However, in the limit of zero softening, it is only guaranteed to yield a correct solution for the linear point-particle perturbation $h^{(1,0)}_{\mu\nu}$, for which classical distribution theory applies and $\psi_\beta$ serves as a delta-net. Beyond that linear perturbation, the limit $\beta\to0$ might not even converge. Moreover, at nonlinear orders the effect of softening is not restricted to be within the softening radius $\beta$ because the softened, $\beta$-dependent perturbations act as sources for higher-order perturbations. But we expect the incurred errors to be small so long as $\beta$ is small compared to the external background length scale $M$.

In Secs.~\ref{sec:instantaneous} and \ref{sec:averaged_forces} we will derive leading-order balance laws compatible with nonlinear matter solutions of the form of Eq.~\eqref{eq:Nonlinearmatter}. For the subleading contributions to the balance laws, which are the necessary quantities for the collections of semi-analytic EMRI-environment models of recent years \cite{Brito:2023pyl,Polcar:2025yto,Li:2025ffh,Dyson:2025env,Keijzer:2026vul,Xu:2026cky}, such as those derived in Sec.~\ref{sec:BalanceTypeC}, we must instead remain within the weakly nonlinear regime introduced next.

\subsubsection{Linear Matter Response}
\label{ssec:linear_response}
Now focusing on the weakly nonlinear or linear-response regime, where both the metric and matter perturbations always remain
small, one may develop a systematic approach to calculating the matter response to higher orders by expanding every quantity jointly in integer powers of the two perturbative parameters. The metric takes the form
\begin{equation}
    {\bf g}_{\mu\nu} := g_{\mu\nu} + \sum_{k+m\geq1}\epsilon^k\lambda^m h_{\mu\nu}^{(k,m)},
\end{equation}
superficially mimicking Eq.~\eqref{eq:self-consistent expansion}. But here the coefficients differ from those in the nonlinear-response formulation because the matter fields $\{\Psi\}$ are expanded
analogously about the unperturbed environmental state: we write $\{\Psi^{(0,1)}\}$ for
that state and $\{\Psi^{(1,1)}\}$ for its linear response to the secondary. For matter
theories whose stress-energy depends nonlinearly on the constituent fields, arranging
integer powers of $\lambda$ at the level of the fields themselves requires a small
amount of extra bookkeeping, but it is still straightforward (see, e.g.,~\cite{Dyson:2025env}).

The environment's stress-energy is also similarly expanded,
\begin{equation}
  \lambda\, T_{(\mathrm{env})}^{\mu\nu}
  =  \sum_{k\geq0,\,m\geq1}\epsilon^{k}\lambda^{m}\,
    T_{(\mathrm{env},k,m)}^{\mu\nu},
  \label{eq:T_expand}
\end{equation}
with the two lowest-order entries
\begin{align}
   T_{(\mathrm{env},0,1)}^{\mu\nu} &= T_{(\mathrm{env})}^{\mu\nu}[g,\{\Psi^{(0,1)}\}],\\
   T_{(\mathrm{env},1,1)}^{\mu\nu} &= \delta^{(1,0)}T_{(\mathrm{env})}^{\mu\nu}[h^{(1,0)},\{\Psi^{(0,1)}\}] \notag\\&\quad+\delta^{(0,1)}T_{(\mathrm{env})}^{\mu\nu}[g,\{\Psi^{(1,1)}\}].
   \label{eq:Tenv11}
\end{align}
The first is simply the unperturbed environment on the background. The second is the
linear correction, which consists of two physically distinct pieces: the direct change
in the stress-energy caused by the secondary's field $h^{(1,0)}_{\mu\nu}$, and the matter's
dynamical response $\{\Psi^{(1,1)}\}$ to that field. For simplicity, we restrict to smooth leading-order matter stress-energies $T^{\mu\nu}_{({\rm env},0,1)}$ that only vary on the time and length scales of the primary black hole, with singularities only arising at higher orders from interaction with the secondary. 

Unlike in the nonlinear-response case, here we do not require a softened potential. Given the perturbative expansion of the matter fields, one can derive an effective point-particle stress-energy tensor for the secondary using matched asymptotic expansions, at least at linear and quadratic orders ($\epsilon$, $\lambda$, $\epsilon^2$, $\epsilon \lambda$, and $\lambda^2$). In a neighbourhood of the secondary, one can first work order by order in $\epsilon$, $\lambda$, and distance from the secondary to derive local solutions for the fields; next, one can carry out a distributional analysis to define a stress-energy tensor that can effectively source these fields~\cite{Upton:2021oxf,Musaeus:2026kgp}. In the vacuum case, if we neglect the secondary's spin, this is the Detweiler stress-energy~\cite{Detweiler:2011tt,Upton:2021oxf},
\begin{equation}\label{eq:TDet}
    \epsilon\, T^{\mu\nu}_{\mathrm{(p)}} = m_{\mathrm{p}}\int_\gamma \tilde u^\mu \tilde u^\nu \frac{\delta^4(x-z(\tilde\tau))}{\sqrt{-\tilde g}}d\tilde\tau,
\end{equation}
where $\tilde g_{\mu\nu} = g_{\mu\nu} + h^{\rm R}_{\mu\nu}$ is a certain smooth effective metric, $z^\alpha(\tilde\tau)$ are the coordinates of the particle's worldline $\gamma$, $\tilde u^\alpha=dz^\alpha/d\tilde\tau$ is the particle's four-velocity, and $\tilde\tau$ is proper time in $\tilde g_{\mu\nu}$. At the orders we consider, Eq.~\eqref{eq:TDet} remains valid in the presence of an environment, given an appropriate definition of the regular field $h^{\rm R}_{\mu\nu}$:
\begin{equation}\label{eq:hR}
    h^{\rm R}_{\mu\nu} = \epsilon h^{({\rm R},1,0)}_{\mu\nu} + \lambda h^{(0,1)}_{\mu\nu} + {\cal O}(\epsilon\lambda,\epsilon^2,\lambda^2).
\end{equation}
Here $h^{({\rm R},1,0)}_{\mu\nu}$ is the Detweiler-Whiting regular field of the particle~\cite{Detweiler:2002mi}, which is a vacuum solution; $h^{(0,1)}_{\mu\nu}$ is the nonvacuum perturbation sourced by the environment's (smooth) stress-energy $T^{\mu\nu}_{({\rm env},0,1)}$. Equation~\eqref{eq:TDet}, with Eq.~\eqref{eq:hR}, follows from the same analysis as in Refs.~\cite{Upton:2021oxf,Musaeus:2026kgp}.

The secondary's stress-energy~\eqref{eq:TDet} can be expanded in the same double-series form as Eq.~\eqref{eq:T_expand},
\begin{equation}
  \epsilon\, T_{(\mathrm{p})}^{\mu\nu}
  = \sum_{k\geq1,\, m\geq0}\epsilon^{k}\lambda^{m}\,
    T_{(\mathrm{p},k,m)}^{\mu\nu},
  \label{eq:Tpp_expand}
\end{equation}
whose leading term is the standard point-particle stress-energy on the Kerr background,
\begin{equation}
T^{\mu\nu}_{(\mathrm{p},1,0)} = m_\mathrm{p}\int_\gamma u^\mu u^\nu\,\frac{\delta^4(x-z(\tau))}{\sqrt{-g}}\,d\tau,
\end{equation}
where $u^\alpha = dz^\alpha/d\tau$ and $\tau$ is proper time in the Kerr background. The term $T^{\alpha\beta}_{(\mathrm{p},1,1)}$ encodes the leading correction to the particle's stress-energy due to the environment in its local zone. 
With this one can expand the Detweiler stress-energy to first order in the environment's metric perturbation $h^{(0,1)}_{\alpha\beta}$ to obtain
\begin{align}\label{eq:Tp11}
T^{\mu\nu}_{(\mathrm{p},1,1)} = m_\mathrm{p}\int_\gamma \frac{1}{2}\left( h^{(0,1)}_{uu} - h^{(0,1)}\right) u^\mu u^\nu\,\frac{\delta^4(x-z(\tau))}{\sqrt{-g}}\,d\tau,
\end{align}
where $h^{(0,1)} \equiv g^{\mu\nu}h^{(0,1)}_{\mu\nu}$ is the trace of the environment's leading-order metric perturbation, and $h^{(0,1)}_{uu} \equiv h^{(0,1)}_{\mu\nu}u^{\mu}u^{\nu}$ arises due to the correction of the secondary's proper time.

Given these stress-energy expansions, we can formulate a rigorous version of the local conservation equation~\eqref{eq:conservation}. An immediate extension of Sec.~IIIB in Ref.~\cite{Musaeus:2026kgp} shows that the correct equation is
\begin{equation}\label{eq:linear-response conservation}
    \epsilon\tilde\nabla_{\mu} T_{(\rm p)}^{\mu\nu} + \lambda \bm{\nabla}_\mu T_{(\env)}^{\mu\nu} = 0
\end{equation}
at the linear and quadratic orders in $\epsilon$ and $\lambda$ that we consider, where $\tilde\nabla_{\mu}$ is the covariant derivative compatible with the effective metric $\tilde g_{\mu\nu}$. In fact, at these orders, we can split the conservation equation into two: 
\begin{equation}\label{eq:particle conservation}
\tilde\nabla_{\mu} T_{(\rm p)}^{\mu\nu}=0
\end{equation}
on the particle's worldline; and 
\begin{equation}
\bm{\nabla}_\mu T_{(\env)}^{\mu\nu} = 0
\end{equation}
everywhere. This follows from the fact that $\tilde\nabla_{\mu} T_{(\rm p)}^{\mu\nu}$ is made up of Dirac-delta terms while $\bm{\nabla}_\mu T_{(\env)}^{\mu\nu}$ cannot contain deltas at these orders because the singular correction to the Christoffel symbols, $\delta\Gamma^\mu_{\nu\gamma}[h^{(1,0)}]$, contains only one derivative of $h^{(1,0)}_{\mu\nu}$, while two are required to generate a delta function (just as for a Newtonian potential). Equation~\eqref{eq:particle conservation} implies that the particle's forced motion in $g_{\mu\nu}$ is geodesic in $\tilde g_{\mu\nu}$ and that its mass $m_{\rm p}$ is constant. 

For compactness we define $T_{(n,m)}^{\mu\nu}$ as the total stress-energy at order $\epsilon^n \lambda^m$,
\begin{equation}
  T_{(n,m)}^{\mu\nu} := T_{(\mathrm{p},n,m)}^{\mu\nu} + T_{(\mathrm{env},n,m)}^{\mu\nu},
  \label{eq:T_total}
\end{equation}
and similarly for metric perturbations. If at any point throughout this work we neglect perturbative indices on a tensorial object, i.e. $h^{(\rm p)}_{\mu\nu}$ or $T_{(\rm env)}^{\mu\nu}$, it should be assumed that the object is the leading term in its perturbative expansion.

\subsection{Green's Functions and Global Conservation Laws}
\label{sec:greensandconserv}
Having constructed the matter response equations, we now introduce the machinery needed to reproduce the Third Law derivation in the relativistic setting. As in the Newtonian derivation, this involves two key Green's-function identities. Unlike the Newtonian case, it requires the construction of global conservation laws, but these too will be related to Newtonian analogues.
\subsubsection{Green's Functions and Metric Perturbations}
Given the stress-energy hierarchy above, we obtain the metric perturbations at each order by inverting the linearised Einstein equations via the retarded Green's function.
The particle and matter fields at leading order are
\begin{align}
h^{(\rm p)}_{\alpha\beta}  &= \int G^{\rm Ret}_{\alpha\beta\alpha'\beta'}T^{\alpha'\beta'}_{(\rm p)}\,dV'_0, \\
h^{(\rm env)}_{\alpha\beta} &= \int G^{\rm Ret}_{\alpha\beta\alpha'\beta'}T^{\alpha'\beta'}_{(\rm env)}\,dV'_0,
\end{align}
in analogy with Eqs.~\eqref{eq:Phi_fl} and \eqref{eq:Phi_p}, where the factor of $8\pi$ has been absorbed into the definition of the Green's function.
Here $T^{\alpha\beta}_{({\rm env})}$ represents $T^{\alpha\beta}_{({\rm env},0,1)}$ in the linear-response case or a solution to Eq.~\eqref{eq:Nonlinearmatter} (with a softened potential) in the nonlinear-response case. $dV_0$ is the covariant background volume element, and the integration runs over the
causal past of each field point due to the presence of the retarded Green's function. 

The construction of the retarded Green's function  requires a choice of gauge; throughout this work we fix all metric perturbations to the Lorenz gauge, $\nabla^{\mu}\bar{h}_{\mu\nu} = 0$, such that they satisfy the linearised equation
\begin{equation}\label{eq:relaxedeins}
{\cal E}_{\alpha\beta}{}^{\mu\nu}\bar h_{\mu\nu}^{(n,m)} = S^{(n,m)}_{\alpha\beta},
\end{equation}
where 
\begin{equation}\label{eq:E operator}
{\cal E}_{\alpha\beta}{}^{\mu\nu}\bar h_{\mu\nu}^{(n,m)} := -\frac{1}{2}(g_\alpha{}^\mu
g_\beta{}^\nu \Box + 2R_\alpha{}^\mu{}_\beta{}^\nu)\bar h_{\mu\nu}^{(n,m)}
\end{equation}
is a relativistic wave operator, with $\Box:=g^{\mu\nu}\nabla_\mu\nabla_\nu$, and $\bar{h}_{\mu\nu}$ is the trace-reversed metric perturbation, $\bar{h}_{\mu\nu} := h_{\mu\nu}- \frac{1}{2}g_{\mu\nu}h^{\alpha}_{\;\alpha}$. The operator ${\cal E}_{\alpha\beta}{}^{\mu\nu}$ is an example of the gauge-fixed linearised Einstein tensor in Eq.~\eqref{eq:Tp soft linear}. The source $S^{(n,m)}_{\alpha\beta}$ includes the particle and environment stress-energy tensors as well as nonlinear variations of the Einstein tensor and any additional source terms; we provide a careful construction of the source in Sec.~\ref{sec:BalanceTypeC}.

We next require the generalisations of the Green's-function identities~\eqref{eq:Newtonian identity 1} and~\eqref{eq:Newtonian identity 2}. First consider Eq.~\eqref{eq:Newtonian identity 2}. The na\"ive  generalisation $G^{\rm Ret}_{\alpha\beta\alpha'\beta'}(x,x') = G^{\rm Ret}_{\alpha'\beta'\alpha\beta}(x',x)$ is plainly false because it involves interchange of $t$ and~$t'$; the interchange $x\leftrightarrow x'$ reverses the time ordering of points. Instead, we must consider the advanced Green's function along with the retarded one. We define the retarded field
$h^{\rm Ret}_{\alpha\beta}$ as the solution to the linearised field equations, sourced
by the particle or environment, with no incoming radiation from the primary's past horizon ($\mathcal{H}^-$) or from past null infinity ($\mathcal{I}^-$). Similarly, the advanced field
$h^{\rm Adv}_{\alpha\beta}$, constructed from the advanced Green's function, is the solution with the same source but with time-reversed propagation; i.e., it has no radiation through the future horizon ($\mathcal{H}^+$) or future null infinity ($\mathcal{I}^+$). The
retarded Green's function satisfies $G^{\rm Ret}_{\alpha\beta\alpha'\beta'}(x,x') = 0$ when $t < t'$ (and more generally, when the field point $x$ is outside the causal future of the source point $x'$), while the advanced Green's function satisfies $G^{\rm
Adv}_{\alpha\beta\alpha'\beta'}(x,x') = 0$ when $t' < t$ (and more generally, when the field point is outside the causal past of the source point). These two Green's functions satisfy a reciprocity relation~\cite{Poisson:2011nh}:
\begin{equation}\label{eq:relativistic identity 1}
    G^{\rm Ret}_{\alpha\beta\alpha'\beta'}(x,x') = G^{\rm Adv}_{\alpha'\beta'\alpha\beta}(x',x).
\end{equation}
This is the relativistic analogue of the Newtonian property~\eqref{eq:Newtonian identity 2}.

Now consider the generalisation of Eq.~\eqref{eq:Newtonian identity 1}. That identity is a result of the translation symmetry of flat space. The generalisations on a curved background are Killing symmetries.  On the Kerr spacetime, which is our primary astrophysical interest, these correspond to the time $\xi^\alpha = (\partial_t)^\alpha$ and the axial Killing vector $\xi^\alpha = (\partial_\phi)^\alpha$. 
For any  Killing vector $\xi^\alpha$ of the background and both Green's
functions, one has
\begin{equation}\label{eq:relativistic identity 2}
{\cal L}_\xi G_{\alpha\beta\alpha'\beta'} = -{\cal L}_{\xi'} G_{\alpha\beta\alpha'\beta'},
\end{equation}
so that Lie differentiation with respect to the field point is equivalent to Lie differentiation
with respect to the source point, up to a sign~\cite{Mino:2003yg}. This is the curved-spacetime analogue of Eq.~\eqref{eq:Newtonian identity 1}. In coordinates, Eq.~\eqref{eq:relativistic identity 2} simply states that $G_{\alpha\beta\alpha'\beta'}$ only depends on $t$ and $t'$ ($\phi$ and $\phi'$) in the combination $t-t'$ ($\phi-\phi'$).

Because the retarded Green's function is not symmetric under $x\leftrightarrow x'$, unlike the Newtonian case, we will also need its decomposition into \emph{symmetric} and  \emph{radiative} parts,
\begin{align}
    h^{\rm Sym/Rad}_{\alpha\beta} &= \frac{1}{2}\left( h^{\rm Ret}_{\alpha\beta}
                               \pm h^{\rm Adv}_{\alpha\beta}\right),
                               \label{eq:hrad_def}
\end{align}
constructed from the two-point functions 
\begin{equation}
    G^{\rm Sym/Rad}_{\alpha\beta\alpha'\beta'} = \frac{1}{2}\left(G^{\rm Ret}_{\alpha\beta\alpha'\beta'}\pm G^{\rm Adv}_{\alpha\beta\alpha'\beta'}\right).
\end{equation}
As $h^{\rm Ret}_{\alpha\beta}$ and $h^{\rm Adv}_{\alpha\beta}$ share the same source,
$h^{\rm Rad}_{\alpha\beta}$ is a homogeneous solution, smooth at the particle, whilst the source of $h^{\rm Sym}_{\alpha\beta}$ remains unchanged, retaining  the singular structure of the retarded field at the secondary.

The radiative field is then time-odd and encodes the outgoing radiation that drives the leading-order orbital evolution in vacuum. By virtue of Eq.~\eqref{eq:relativistic identity 1}, $G^{\rm Rad}_{\alpha\beta\alpha'\beta'}$ is anti-symmetric,
\begin{equation}
    G^{\rm Rad}_{\alpha\beta\alpha'\beta'}(x,x') = -G^{\rm Rad}_{\alpha'\beta'\alpha\beta}(x',x),
\end{equation}
whilst naturally,
$G^{\rm Sym}_{\alpha\beta\alpha'\beta'}$ is symmetric,
\begin{equation}
    G^{\rm Sym}_{\alpha\beta\alpha'\beta'}(x,x') = G^{\rm Sym}_{\alpha'\beta'\alpha\beta}(x',x),
\end{equation}
and Eq.~\eqref{eq:relativistic identity 2} then implies
\begin{align}
{\cal L}_\xi G^{\rm Rad}_{\alpha\beta\alpha'\beta'} &= +{\cal L}_{\xi'} G^{\rm Rad}_{\alpha'\beta'\alpha\beta}. \label{Lie GRad}
\end{align}
with
\begin{align}
{\cal L}_\xi G^{\rm Sym}_{\alpha\beta\alpha'\beta'} &= -{\cal L}_{\xi'} G^{\rm Sym}_{\alpha'\beta'\alpha\beta}. \label{Lie GSym}
\end{align}

\subsubsection{Conserved Charges}
\label{sec:conservation}
In the Newtonian case, the Third Law was formulated in terms of the total force on the matter. However, we could instead consider what that total force represents: the rate of change of the matter's total linear momentum. We cannot simply sum forces on a curved spacetime, but we can define quantities analogous to total linear momentum and calculate their rates of change. Our relativistic Third Law will then be a relationship between the rate of change of the matter's generalised momenta and the rate of change of the particle's. 

These generalised momenta, which we refer to as conserved charges, are associated with the Killing vectors of the background spacetime, in the same way Newtonian linear momentum is associated with the translation symmetry of Euclidean space.  For a geodesic of the background, the momenta conjugate to these symmetries, $m_\mathrm{p} u_\alpha \xi^\alpha$, are constants of the motion, and it is these that we generalise below to the charges $P_\xi$ which we use to evolve our orbits. For the point particle these are directly related to the specific orbital energy and $z$ component of the specific orbital angular momentum,
\begin{equation}
\mathcal{E}^{(\mathrm{p})} = -u_t  \qquad \mathcal{L}^{(\mathrm{p})} _z = u_\phi ,
\end{equation}
where the sign convention ensures $\mathcal{E}^{(\mathrm{p})}>0$. These are the constants of motion that determine the adiabatic inspiral for equatorial orbits~\cite{Mino:1996nf, Barack:2018yvs}. A third constant of geodesic motion in Kerr, the Carter constant $\mathcal{Q}$, is associated with a
rank-two Killing tensor rather than a Killing vector. The results we present here, while valid for generic, inclined orbits, only provide prescriptions for the evolution of orbital elements directly associated with Killing vectors, meaning our prescription is only complete for equatorial orbits and for environments that preserve them. Further work will be required to extend the Third Law to the evolution of the Carter constant.

Our strategy for energy and angular momentum is as follows. In a perturbed spacetime ${\bf g}_{\alpha\beta} = g^{\rm Kerr}_{\alpha\beta} + h_{\alpha\beta}$, the charge $P_\xi$ defined below is no longer conserved; it changes because $\xi^\alpha$ is a Killing vector of the background only, not of the full metric. The rate of change is generated by $T^{\alpha\beta}\mathcal{L}_\xi h_{\alpha\beta}$, which couples the stress-energy to the metric perturbation, and it is this coupling, expanded order by order in $(\epsilon, \lambda)$, that produces the relativistic power and torque. In this section we derive the general formulae for $\dot{P}_\xi$ that will be applied in Secs.~\ref{sec:instantaneous} and~\ref{sec:averaged_forces}; see Sec.~IV of Ref.~\cite{Rahman:2026qho} for recent, similar manipulations.

For simplicity, as in the opening of Sec.~\ref{sec:TwoParamframwork}, we derive the basic results while imagining all fields are smooth; we explain the applicability to our case at the end of the section. Consider a generic spacetime metric ${\bf g}_{\alpha\beta}$, a stress-energy tensor $T^{\alpha\beta}$ conserved with respect to ${\bf g}_{\alpha\beta}$, a vector field $\xi^\alpha$, and a family of spacelike hypersurfaces $\Sigma_t$ of constant time $t$. Associated with these, we define a charge $P_\xi$,
\begin{equation}\label{Q def}
P_\xi(\Sigma_t) = \int_{\Sigma_t} T^\alpha_{\ \beta}\, \xi^\beta \, d\Sigma_\alpha,
\end{equation}
where
\begin{equation}
d\Sigma_\alpha = -n_\alpha \, d\Sigma = -n_\alpha \sqrt{-{\bf g}}\,d^3x
\end{equation}
is the surface element from ${\bf g}_{\alpha\beta}$. Here $n_\alpha = - \partial_\alpha t$ is a future-directed normal vector to $\Sigma_t$. 
With this choice, evaluating the leading-order contribution to Eq.~\eqref{Q def} from the point-particle stress-energy gives
\begin{equation}
P_t^{(\mathrm{p})} = -m_\mathrm{p} \mathcal{E}^{(\mathrm{p})}, \qquad P^{(\mathrm{p})}_\phi = m_\mathrm{p} \mathcal{L}^{(\mathrm{p})}_z.
\end{equation}
At the first subleading order, which is required for consistent orbital evolution, these are corrected by the metric perturbation from the environment.

\begin{figure}[tb!]
  \centering
\includegraphics[width=0.95\columnwidth]{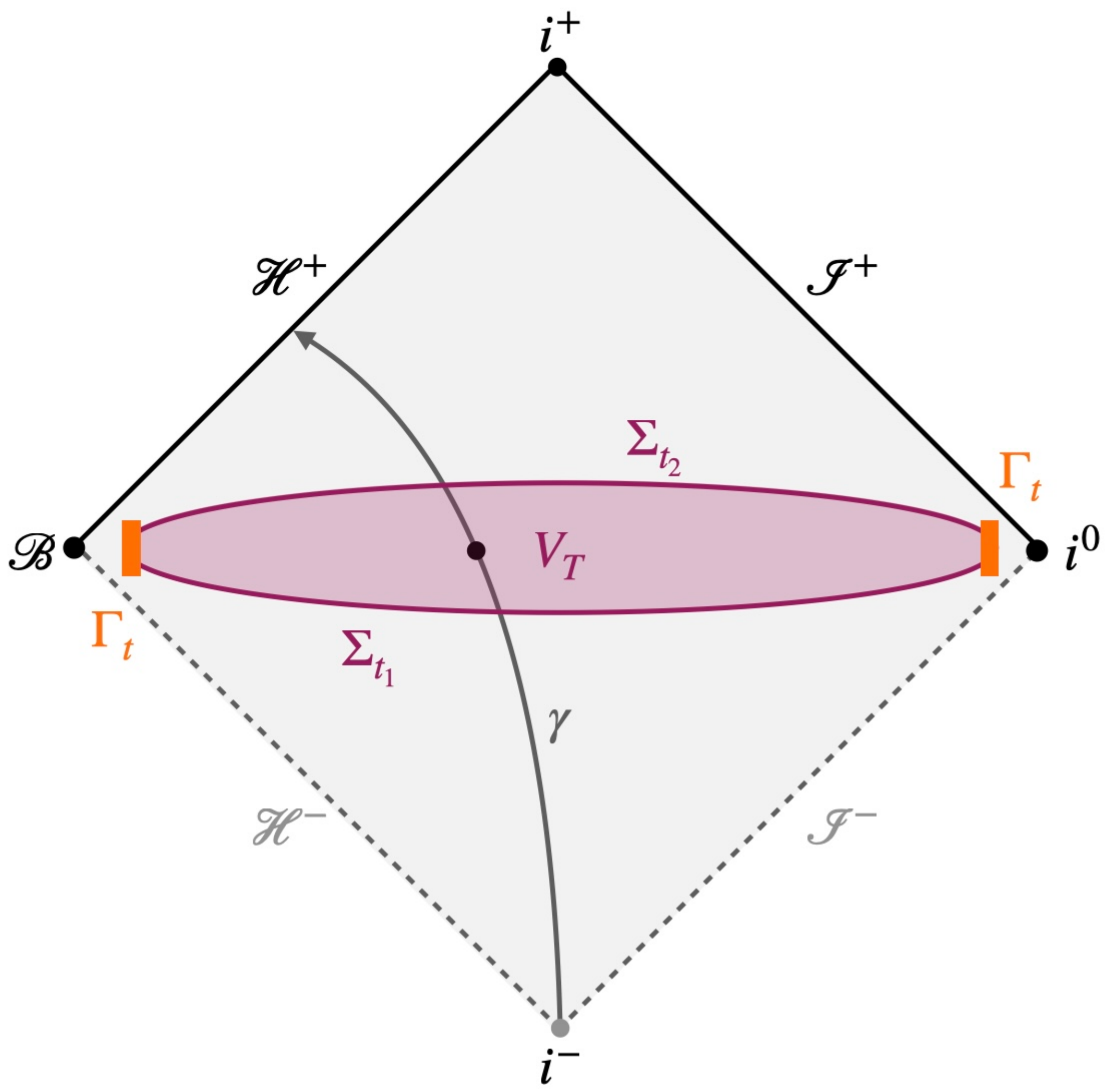}
  \caption{Penrose diagram of the black hole exterior. The constant-$t$
    hypersurfaces $\Sigma_{t_1}$ and $\Sigma_{t_2}$ run between two timelike surfaces, one ``close'' to the bifurcation 2-sphere and the other ``close'' to spatial infinity, bounding the volume
    $V_T$ (shaded). The orbiting secondary travels on the worldline $\gamma$ from $i^-$, passes through $V_T$ at $x$, and crosses
    $\mathcal{H}^+$.}
  \label{fig:penrose-hypersurface}
\end{figure}

We are interested in the change in the full quantity $P_\xi$ between two slices. For concreteness, we choose a reference time $t$ and initial and final times
\begin{equation}
t_1 = t-T \qquad\text{and}\qquad t_2 = t+T
\end{equation}
for some $T > 0$; this choice will allow us to discuss the average rate of change at time $t$. We first consider the change
\begin{equation}
\Delta_T P_\xi := P_\xi(\Sigma_{t_2}) - P_\xi(\Sigma_{t_1})
\end{equation}
between the charges on the two slices. We define a volume $V_T$ with boundary $\partial V_T = \Sigma_{t_2}\cup \Sigma_{t_1}\cup \Gamma_T$, where $\Gamma_T$ is a timelike surface defined as in Fig.~\ref{fig:penrose-hypersurface}. Applying Stokes' theorem, we then find
\begin{align}
\Delta_T P_\xi
  &= \int_{V_T} {\bm \nabla}_\alpha (T^{\alpha\beta} \xi_\beta)\,dV
     - \int_{\Gamma_T} T^\alpha_{\ \beta} \xi^\beta\,d\Gamma_\alpha \nonumber \\
  &= \frac{1}{2}\int_{V_T} T^{\alpha\beta}{\cal L}_\xi {\bf g}_{\alpha\beta}\,dV
     - \int_{\Gamma_T} T^\alpha_{\ \beta} \xi^\beta\,d\Gamma_\alpha,
     \label{Delta Q v1}
\end{align}
where we have used ${\bm \nabla}_\beta T^{\alpha\beta} = 0$, $d\Gamma_{\alpha}$ is the outward-directed surface element on $\Gamma_T$, and $dV = dt\,d\Sigma$ is the 4-dimensional volume element.

Taking the limit of two surfaces infinitesimally close, we can now write the instantaneous rate of change as
\begin{equation}\label{eq:fullQdot}
\frac{dP_\xi}{dt} = \frac{1}{2}\int_{\Sigma_t} T^{\alpha\beta}{\cal L}_\xi h_{\alpha\beta}\,d\Sigma - \int_{\partial\Sigma_t}T^{\alpha\beta}\xi_{\beta}r_{\alpha}\, dS,
\end{equation}
where we now used the fact that $\xi^\alpha$ is a Killing vector of the background. Here $r_\alpha$ is the outward unit normal to $\partial\Sigma_t$ tangent to $\Sigma_t$, $dS$
is the induced area element, and $\partial\Sigma_t$ is taken to the surface where the
slices meet (the bifurcation sphere $\mathcal{B}$ in the stationary limit) and to $i^{0}$. If $\xi^\alpha$ were a Killing vector of the full spacetime, then the integral over $\Sigma_t$ would vanish. Any change in $P_\xi$ would then arise from the second term in Eq.~\eqref{eq:fullQdot}: the matter flux through the boundary.

For stress-energies of spatially compact support away from the boundaries, such as that of the point particle, the boundary contribution vanishes identically, and the averaged rate of change reduces to
\begin{equation}\label{avg Qdot}
\left\langle \frac{dP_\xi}{dt}\right\rangle
  = \lim_{T\to\infty}\frac{\Delta_T P_\xi}{2T}
  = \lim_{T\to\infty}\frac{1}{4T}\int_{V_T} T^{\alpha\beta}{\cal L}_\xi h_{\alpha\beta}\,dV.
\end{equation}
This might not be the case for the environment, which could contain a nonzero matter flux.

Finally, we address the applicability of these results in our case, where the metric and stress-energy tensor are singular on the worldline $\gamma$. If we apply the formulas na\"{\i}vely, nearly all the expressions become ill-defined due to non-integrable products of singularities. We avoid this problem in different ways depending on whether we are studying the linear-response or nonlinear-response regimes. In the linear-response case, we note that our derivation of the Third Law is initially only concerned with the leading-order interaction between the particle and the environment. That interaction manifests itself in cross terms of order $\epsilon\times\lambda$ in the above derivation (though we bootstrap from these results to obtain $\epsilon^2\times\lambda$ terms in Sec.~\ref{sec:BalanceTypeC}). Since the environment's leading-order stress-energy tensor $T^{\alpha\beta}_{({\rm env},0,1)}$ and field $h^{(\env, 0,1)}_{\alpha\beta}$ are smooth, the cross terms are all well defined. Every equality ``$X = Y$'' in the above derivation should then be understood as ``order-$\epsilon\lambda$ terms in $X$ are equal to order-$\epsilon\lambda$ terms in $Y$''. This does not work in the nonlinear-response case because even the leading-order $T^{\alpha\beta}_{(\rm env)}$, satisfying  Eq.~\eqref{eq:Nonlinearmatter}, could be non-integrable. However, in this case all expressions are made well defined through our use of a softened particle potential---though we reiterate our warning that such an approach is an approximation that must be carefully considered in any practical implementation and might not converge in the limit of zero softening.

\section{The Relativistic Third Law}
\label{sec:BalanceFirst}
In this section we derive the instantaneous balance law at leading order, together with its orbit-averaged form. Importantly, all results at $\mathcal{O}(\epsilon\lambda)$ apply equally to the nonlinear-response formulation of Sec.~\ref{sec:nonlinear_response} and to the leading-order linear response equations of Sec.~\ref{ssec:linear_response}.
\subsection{Instantaneous Balance}
\label{sec:instantaneous}
Starting from Eq.~\eqref{eq:fullQdot}, we consider the instantaneous rate of change of the particle's orbital
charge due to $\mathcal{O}(\epsilon\lambda)$ cross-coupling
between the particle and matter stress-energies. Defining the particle's contribution to the charge~\eqref{Q def} with the replacement $T^{\alpha\beta}\to \epsilon T^{\alpha\beta}_{(\mathrm{p})}$, we can ascertain that the order-$\epsilon\lambda$ term in Eq.~\eqref{eq:fullQdot} reads
\begin{equation}\label{Qdot_inst}
\frac{dP^{(\rm p)}_\xi}{dt} = \frac{1}{2}\int_{\Sigma_t} T^{\alpha\beta}_{(\rm p)}{\cal L}_\xi h_{\alpha\beta}\,d\Sigma_0,
\end{equation}
where $d\Sigma_0$ is the background Kerr surface element and the boundary term now vanishes due to the  point-particle stress-energy tensor having zero support on the boundaries. We again emphasise that this equality should be interpreted as an equality between leading cross terms.

It might not be obvious that Eq.~\eqref{Qdot_inst} holds, for the following reason: if we replace $T^{\alpha\beta}$ with $T^{\alpha\beta}_{(\rm p)}$ in the derivation of Eq.~\eqref{eq:fullQdot}, we seem to require $\bm{\nabla}_\beta T^{\alpha\beta}_{(\rm p)}=0$ in Eq.~\eqref{Delta Q v1}. In reality, the conservation equation is instead Eq.~\eqref{eq:conservation} with a softened particle or Eq.~\eqref{eq:linear-response conservation}. However, we do have $\epsilon\bm{\nabla}_\beta T^{\alpha\beta}_{(\rm p)} = {\cal O}(\epsilon\lambda^2)$ in the sense described below Eq.~\eqref{avg Qdot} (i.e., only keeping leading cross terms). We can neglect this violation in our leading-cross-term analysis.

Now, we first isolate the $\mathcal{O}(\epsilon\lambda)$ cross-term in Eq.~\eqref{Qdot_inst} and substitute the retarded representation of $h^{(\env)}_{\alpha\beta}$, leading to
\begin{equation}\label{eq:standardforce}
     \frac{dP^{(\rm p)}_\xi}{dt}
     = \frac{1}{2}\int_{\Sigma_{t}} 
        T_{(\rm p)}^{\alpha\beta}\,{\cal L}_\xi
       h^{\rm (env,Ret)}_{\alpha\beta}d\Sigma_0.
\end{equation}
In a more symmetrical form analogous to Eq.~\eqref{eq:Fp_step2},
\begin{equation}
    \frac{dP^{\mathrm{(\mathrm{p})}}_{\xi}}{dt} = \frac{1}{2}\int_{-\infty}^t\! dt'\iint d\Sigma_0\, d\Sigma_0'\, T^{\alpha\beta}_{(\mathrm{p})}\,{\cal L}_\xi G^{\rm (Ret)}_{\alpha\beta\alpha'\beta'}T^{\alpha'\beta'}_{(\mathrm{env})},
\end{equation}
where the spatial integrals are over $\Sigma_t$ and $\Sigma_{t'}$, and we have truncated the integral over $t'$ using the causal relationship enforced by the Green's function.

We next use the Green's-function identities~\eqref{eq:relativistic identity 1} and~\eqref{eq:relativistic identity 2} to rewrite this rate of change as
\begin{align}
\frac{dP^{\mathrm{(\mathrm{p})}}_{\xi}}{dt}
  &= -\frac{1}{2}\int_{-\infty}^t\! dt'\iint d\Sigma_0\, d\Sigma_0'\,T^{\alpha'\beta'}_{(\rm env)}\,
     \mathcal{L}_{{\xi'}}G^{\mathrm{Adv}}_{\alpha'\beta'\alpha\beta}\,
     T^{\alpha\beta}_{(\rm p)},
\end{align}
moving the Lie derivative from the point on the particle to the point in the environment. To put this in the form~\eqref{eq:Newton2.9}, we define an effective stress-energy tensor,
\begin{align}
  T^{\alpha\beta}_{({\rm p} | t^{*})} :=  T^{\alpha\beta}_{(\rm p)}(t,x^i)\;\delta(t-t^{*}),
\end{align}
which localises the point-particle source to the time $t^*$.
Subsequently we define the time-localised point particle metric perturbation,
\begin{equation}\label{eq:localised hAdv}
    h^{{\rm(p,Adv}|t)}_{\alpha'\beta'} = \int dV_0'' \,G^{\rm Adv}_{\alpha'\beta'\alpha''\beta''}T^{\alpha''\beta''}_{({\rm p}|t)},
\end{equation}
where $h^{(\mathrm{p,Adv}|t)}_{\alpha'\beta'}$ is the advanced metric perturbation whose source is localised to the slice $\Sigma_t$ on which the rate of change of the charge is evaluated; notably, this field is sourced by a non-conserved stress-energy tensor. This gives the final expression,
\begin{equation}\label{eq:BalanceLaw}
  \frac{dP^{(\mathrm{p})}_{\xi}}{dt}
  = - \frac{1}{2}\int_{-\infty}^{t} dt'\!\int_{\Sigma_{t'}}
    T^{\alpha'\beta'}_{(\mathrm{env})}\;\mathcal{L}_{\xi'}
       h^{(\mathrm{p,Adv}|t)}_{\alpha'\beta'}\,d\Sigma_0',
\end{equation}
which is the relativistic extension of Newton's Third Law to charges associated with approximate Killing symmetries in curved spacetime.

We note that integrating the localised perturbation over all time recovers the full perturbation,
\begin{equation}
    \int_{-\infty}^{\infty} h^{(\cdot|t)}_{\alpha'\beta'}(t',\bm{x}')\,dt
    = h^{(\cdot)}_{\alpha'\beta'}(t',\bm{x}').
\end{equation}
Importantly, $h^{(\mathrm{p,Adv}|t)}_{\alpha\beta}$ is sourced by a
non-conserved stress-energy tensor; the solution is purely vacuum except at a single point in spacetime. We expect such a perturbation to be computed as a vacuum solution to the relaxed Einstein equations, subject to conservation-violating jump
initial conditions; demonstrating this practically will be an important direction of future work.

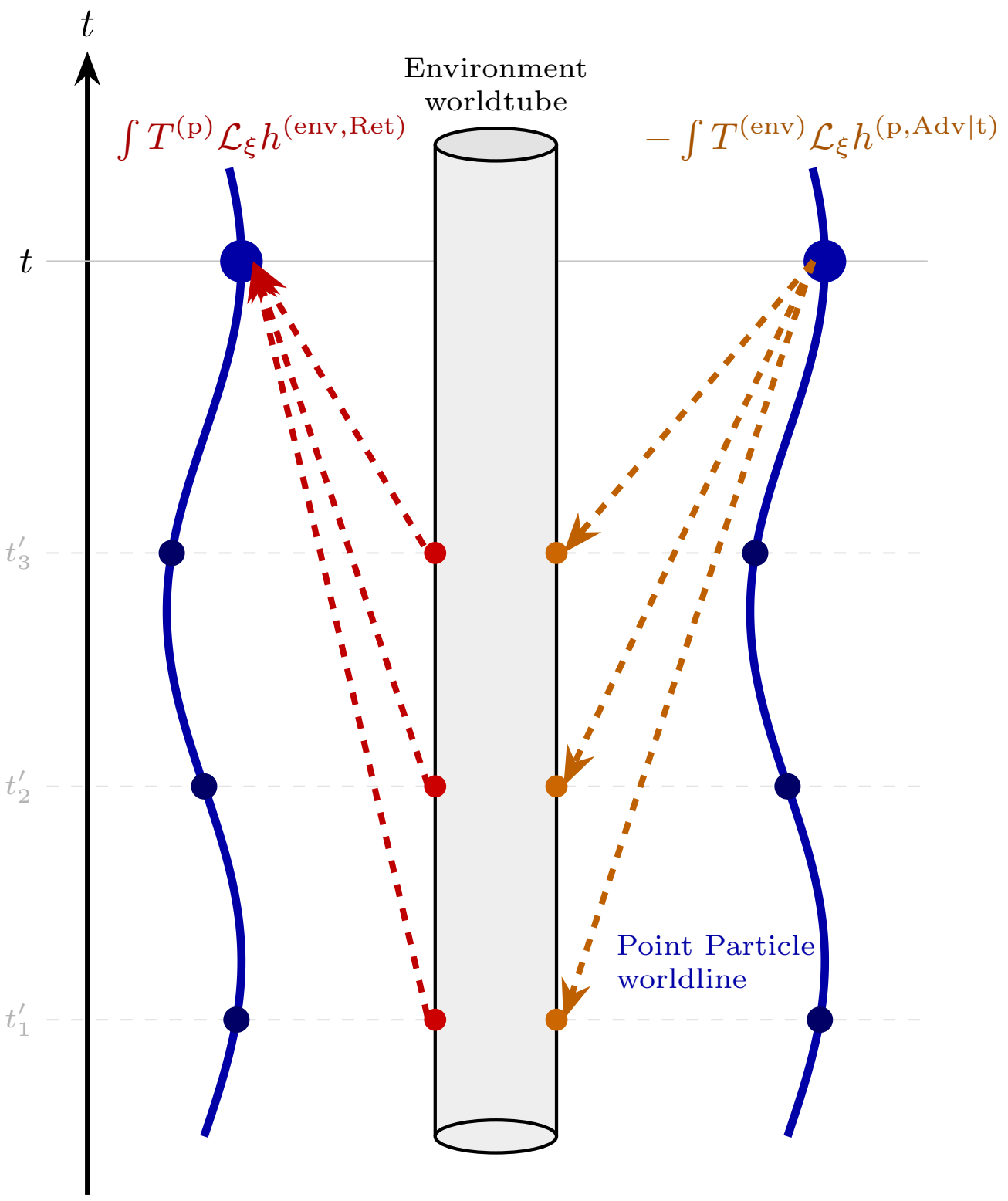
\begin{figure}

\begin{tikzpicture}[>=Stealth, font=\normalsize]

\def\tCur{8.0}

\def\FX{3.5}
\def\FR{0.52}

\def\WLRcenter{6.0}
\def\WLamp{0.32}
\def\WLperiod{6.0}
\def\WLphase{0.5}

\def\WLLcenter{1.0}

\def\Eya{1.5}
\def\Eyb{3.5}
\def\Eyc{5.5}

\pgfmathsetmacro{\Exa}{\WLRcenter+\WLamp*sin(360*(\Eya-\WLphase)/\WLperiod)}
\pgfmathsetmacro{\Exb}{\WLRcenter+\WLamp*sin(360*(\Eyb-\WLphase)/\WLperiod)}
\pgfmathsetmacro{\Exc}{\WLRcenter+\WLamp*sin(360*(\Eyc-\WLphase)/\WLperiod)}
\pgfmathsetmacro{\Ext}{\WLRcenter+\WLamp*sin(360*(\tCur-\WLphase)/\WLperiod)}

\pgfmathsetmacro{\ExLa}{\WLLcenter+\WLamp*sin(360*(\Eya-\WLphase)/\WLperiod)}
\pgfmathsetmacro{\ExLb}{\WLLcenter+\WLamp*sin(360*(\Eyb-\WLphase)/\WLperiod)}
\pgfmathsetmacro{\ExLc}{\WLLcenter+\WLamp*sin(360*(\Eyc-\WLphase)/\WLperiod)}
\pgfmathsetmacro{\ExtL}{\WLLcenter+\WLamp*sin(360*(\tCur-\WLphase)/\WLperiod)}

\draw[->, line width=1.2pt] (0,0) -- (0,9.8)
  node[above, font=\normalsize] {$t$};

\draw[gray!45, thin] (-0.35,\tCur) -- (7.2,\tCur);
\node[left, font=\small] at (-0.35,\tCur) {$t$};
\foreach \y/\lbl in {\Eya/{t'_1}, \Eyb/{t'_2}, \Eyc/{t'_3}} {
  \draw[gray!22, thin, dashed] (-0.35,\y) -- (7.2,\y);
  \node[left, gray!60, font=\scriptsize] at (-0.35,\y) {$\lbl$};
}

\fill[gray!13] ({\FX-\FR},0.5) rectangle ({\FX+\FR},9.0);
\draw[thick]   ({\FX-\FR},0.5) -- ({\FX-\FR},9.0);
\draw[thick]   ({\FX+\FR},0.5) -- ({\FX+\FR},9.0);
\draw[thick, fill=gray!22] (\FX,9.0) ellipse (0.52 and 0.14);
\draw[thick, fill=gray!13] (\FX, 0.5) ellipse (0.52 and 0.14);
\node[above, align=center, font=\scriptsize] at (\FX,9.18)
  {Environment\\worldtube};

\draw[line width=2pt, blue!65!black]
  plot[domain=0.5:8.8, samples=150, smooth]
  ({\WLRcenter+\WLamp*sin(360*(\x-\WLphase)/\WLperiod)}, {\x});
\node[right, blue!65!black, align=left, font=\scriptsize]
  at ({\WLRcenter+\WLamp-1.9}, 2.0) {Point Particle\\worldline };

\draw[line width=2pt, blue!65!black]
  plot[domain=0.5:8.8, samples=150, smooth]
  ({\WLLcenter+\WLamp*sin(360*(\x-\WLphase)/\WLperiod)}, {\x});

\coordinate (Pa)  at (\Exa,  \Eya);
\coordinate (Pb)  at (\Exb,  \Eyb);
\coordinate (Pc)  at (\Exc,  \Eyc);
\coordinate (P)   at (\Ext,  \tCur);

\coordinate (PLa) at (\ExLa, \Eya);
\coordinate (PLb) at (\ExLb, \Eyb);
\coordinate (PLc) at (\ExLc, \Eyc);
\coordinate (PL)  at (\ExtL, \tCur);

\coordinate (Fca)  at ({\FX+\FR}, \Eya);
\coordinate (Fcb)  at ({\FX+\FR}, \Eyb);
\coordinate (Fcc)  at ({\FX+\FR}, \Eyc);

\coordinate (FLca) at ({\FX-\FR}, \Eya);
\coordinate (FLcb) at ({\FX-\FR}, \Eyb);
\coordinate (FLcc) at ({\FX-\FR}, \Eyc);

\filldraw[blue!40!black] (Pa) circle (3pt);
\filldraw[blue!40!black] (Pb) circle (3pt);
\filldraw[blue!40!black] (Pc) circle (3pt);
\filldraw[blue!65!black] (P)  circle (5pt);

\filldraw[blue!40!black] (PLa) circle (3pt);
\filldraw[blue!40!black] (PLb) circle (3pt);
\filldraw[blue!40!black] (PLc) circle (3pt);
\filldraw[blue!65!black] (PL)  circle (5pt);

\draw[->, dashed, line width=1.5pt, orange!75!black]
  ($(P)+(-0.10,0)$) -- ($(Fca)+(0.06,0)$);
\draw[->, dashed, line width=1.5pt, orange!75!black]
  ($(P)+(-0.10,0)$) -- ($(Fcb)+(0.06,0)$);
\draw[->, dashed, line width=1.5pt, orange!75!black]
  ($(P)+(-0.10,0)$) -- ($(Fcc)+(0.06,0)$);

\filldraw[orange!80!black] (Fca) circle (2.5pt);
\filldraw[orange!80!black] (Fcb) circle (2.5pt);
\filldraw[orange!80!black] (Fcc) circle (2.5pt);

\draw[<-, dashed, line width=1.5pt, red!75!black]
  ($(PL)+(0.10,0)$) -- ($(FLca)+(-0.06,0)$);
\draw[<-, dashed, line width=1.5pt, red!75!black]
  ($(PL)+(0.10,0)$) -- ($(FLcb)+(-0.06,0)$);
\draw[<-, dashed, line width=1.5pt, red!75!black]
  ($(PL)+(0.10,0)$) -- ($(FLcc)+(-0.06,0)$);

\filldraw[red!80!black] (FLca) circle (2.5pt);
\filldraw[red!80!black] (FLcb) circle (2.5pt);
\filldraw[red!80!black] (FLcc) circle (2.5pt);

\node[above, orange!65!black, font=\small] at (6.3, 8.75)
  {$-\int T^{\rm (env)}\mathcal{L}_{\xi}h^{(\rm p,Adv|t)}$};
\node[above, red!65!black, font=\small] at (1.5, 8.75)
  {$\int T^{(\mathrm{p})} \mathcal{L}_{\xi}h^{(\env,\rm Ret)}$};

\end{tikzpicture}
\caption{Visual representation of how the balance law takes the force on the particle from the environment's retarded field (left side of diagram) and relates it to the environment's perspective of how it will interact with the particle's field in the future (right side of diagram), showing explicitly the causal form of Newton's Third Law in curved space from the view of distinct time slices.}
\label{fig:Worldline}
\end{figure}

We now briefly interpret Eq.~\eqref{eq:BalanceLaw} through the diagrammatic
representation in Fig.~\ref{fig:Worldline}. Newton's Third Law at its core is a statement
that the force exerted on object $A$ by object $B$ is equal and opposite to the force exerted on $B$ by $A$ at that instant. In its na\"{\i}ve form this is an instantaneous, acausal relation and breaks down even in flat spacetime once retardation effects are accounted for.

In the relativistic setting the interaction occurs in the following way instead: the environment generates a retarded gravitational field that propagates forward in time, and at time $t$ this results in a field sourced by all prior times $t' < t$. The derivative of this field evaluated on the worldline and contracted with the four-velocity at the measurement time then gives the force on the secondary. This is precisely the force appearing on the right-hand side of Eq.~\eqref{eq:standardforce} and illustrated by the red arrows in Fig.~\ref{fig:Worldline}.

The balance law Eq.~\eqref{eq:BalanceLaw} is then the relativistic, causal form of Newton's Third Law: it expresses the same rate of change from the perspective of the environment. To identify the reaction force in the environment's frame at a fixed instant $t$, one must integrate over the environment's entire past history $t' < t$, collecting every way in which the environment at earlier times interacts with the particle at time $t$. This is precisely what the localised advanced field $h^{(\mathrm{p,Adv}|t)}_{\alpha\beta}$ encodes: in the time-reversed picture, it is the field emitted by the particle at time $t$ and received by the environment at the earlier time $t'$. The right-hand side of Eq.~\eqref{eq:BalanceLaw} therefore integrates the environment's past stress-energy against this localised advanced field, as illustrated by the orange arrows in Fig.~\ref{fig:Worldline}.

The localisation of the particle source to the single time $t$ can also be understood from this picture. It is constructed such that the past instances of the environment, at all times $t'<t$, receive the advanced field from only a single future time slice, i.e. the slice on which we measure the force. Without this localisation, the integral would account for the full accumulated interaction between the environment and the particle along the entire worldline. Although it may seem strange, it is precisely this localisation that makes Eq.~\eqref{eq:BalanceLaw} the correct expression for
$dP^{(\mathrm{p})}_\xi/dt$ at a given instant.

\subsection{Averaged Balance}
\label{sec:averaged_forces}
Having derived the instantaneous-force formula in Sec.~\ref{sec:instantaneous}, we now
turn to the orbit-averaged rates of change, for which Eq.~\eqref{avg Qdot} reduces to
\begin{equation}\label{avg Qdot pert}
\left\langle \frac{dP^{(\rm p)}_\xi}{dt}\right\rangle
  = \lim_{T\to\infty}\frac{1}{4T}\int_{V_T} T^{\alpha\beta}_{(\rm p)}{\cal L}_\xi h_{\alpha\beta}\,dV_0,
\end{equation}
where, as before, $h_{\alpha\beta}$ is the sum of all perturbations. Keeping only the leading cross term and writing the
metric perturbation in terms of the retarded Green's function, we then have
\begin{align}
\left\langle \frac{dP^{(\rm p)}_\xi}{dt}\right\rangle
  &= \lim_{T,T'\to\infty}\frac{1}{4T}
    \int_{V_T}\int_{V_{T'}}
     T_{(\rm p)}^{\alpha\beta}\,\,\notag \\
     &\qquad\qquad
    \times {\cal L}_\xi G^{\rm Ret}_{\alpha\beta\alpha'\beta'} T_{(\rm env)}^{\alpha'\beta'}\,dV_0\,dV_0' \\[5pt]
  &= -\lim_{T,T'\to\infty}\frac{1}{4T}
    \int_{V_T}\int_{V_{T'}}
     T_{(\rm env)}^{\alpha'\beta'}\,\notag \\
    &\qquad\qquad
    \times{\cal L}_{\xi'} G^{\rm Adv}_{\alpha'\beta'\alpha\beta}\,
     T_{(\rm p)}^{\alpha\beta}\,dV_0\,dV_0',
    \label{<Qdot>}
\end{align}
which we can rewrite as
\begin{equation}\label{eq:averaged 3rd law v1}
    \left\langle \frac{dP^{(\rm p)}_\xi}{dt}\right\rangle
 = -\lim_{T\to\infty}\frac{1}{4T}
    \int_{V_T}
     T_{(\rm env)}^{\alpha\beta}\,{\cal L}_{\xi}\, h^{(\rm p,\,Adv)}_{\alpha\beta}\,dV_0.
\end{equation}

Making use of the identity $h^{\rm Adv}_{\alpha\beta} = h^{\rm Ret}_{\alpha\beta}-2h^{\rm Rad}_{\alpha\beta}$, and defining the average power (or torque) functional on Kerr as
\begin{equation}\label{eq:work_functional}
    \mathcal{W}_{\xi}[T,h]:= \lim_{T\to\infty}\frac{1}{4T}\int_{V_T}
     T^{\alpha\beta}\,{\cal L}_{\xi}h_{\alpha\beta} dV_0 ,
\end{equation}
we arrive at the orbit-averaged relativistic Third Law:
\begin{equation}\label{eq:AveragedThirdLaw}
\left\langle \frac{dP^{(\rm p)}_\xi}{dt}\right\rangle
  = -\left\langle F^{(\rm env)}_\xi\right\rangle
    - \left\langle{\cal F}_\xi^{(\env\times{\rm p})}\right\rangle,
\end{equation}
where 
\begin{align}
    \left\langle F^{(\rm env)}_\xi\right\rangle
    = \mathcal{W}_{\xi}[T_{(\env)},h^{(\rm p,Ret)}]\label{eq:AveragedThirdLaw_Fluxes}
\end{align}
is the average rate of work done (or torque exerted) on the environment by the secondary, and ${\cal F}_\xi^{(\env\times{\rm p})}$ collects all energy (or angular momentum) carried to
infinity and down the horizon by gravitational waves sourced jointly by the secondary
and the matter at $\mathcal{O}(\epsilon\lambda)$,
\begin{equation}\label{eq:fluxleading}
   \left\langle {\cal F}_\xi^{(\env\times{\rm p)}}\right\rangle
    = -2\mathcal{W}_{\xi}[T_{(\env)},h^{(\rm p,Rad)}].
\end{equation}

To see why Eq.~\eqref{eq:fluxleading} is the total order-$\epsilon\lambda$ flux, first note the identity~\eqref{Lie GRad} allows us to write it in the symmetrical form
\begin{align}
       \left\langle {\cal F}_\xi^{(\env\times{\rm p})}\right\rangle
    = &-\mathcal{W}_{\xi}[T_{(\env)},h^{(\rm p,Rad)}]\notag\\
    &-\mathcal{W}_{\xi}[T_{(\rm p)},h^{(\rm env,Rad)}].
\end{align}
We can combine this with the $\epsilon^2$ and $\lambda^2$ fluxes, 
\begin{equation}
\left\langle {\cal F}_\xi^{({\rm p})}\right\rangle
    = -\mathcal{W}_{\xi}[T_{(\rm p)},h^{(\rm p,Rad)}]
\end{equation}      
and 
\begin{equation}
    \left\langle {\cal F}_\xi^{({\rm env})}\right\rangle
    = -\mathcal{W}_{\xi}[T_{(\rm env)},h^{(\rm env,Rad)}],
\end{equation}
to write the sum of $\epsilon^2$, $\lambda^2$, and $\epsilon\lambda$ fluxes as
\begin{equation}\label{eq:total flux}
    \left\langle {\cal F}_\xi\right\rangle
    = -\mathcal{W}_{\xi}[T,h^{(\rm Rad)}],
\end{equation}
where $T^{\alpha\beta}$ and $h^{(\rm Rad)}_{\alpha\beta}$ are the sums of the particle's and environment's stress-energies and radiative fields, respectively. 

Next, the volume integral in Eq.~\eqref{eq:total flux} can be expressed in the standard form of an asymptotic flux by exploiting the factorised (mode-sum) representation of
$G^{\rm Rad}_{\alpha\beta\alpha'\beta'}$, which decomposes into a sum of products of ingoing and outgoing wave solutions~\cite{Galtsov:1982hwm,Sago:2005fn,Akcay:2019bvk}. Dividing each flux into a sum of the fluxes out to infinity and down the horizon, as in ${\cal F}_{\xi}={\cal F}^\infty_{\xi}+{\cal F}^{\cal H}_{\xi}$, we can write~\cite{Poisson:2004cw,Compere:2019gft}
\begin{align}
    \left\langle {\cal F}^\infty_{\xi}\right\rangle &= -\frac{1}{32\pi}\lim_{r\to\infty}\left\langle \int {\cal L}_t h^{AB}{\cal L}_\xi h_{AB}\, dS_0\right\rangle,\label{eq:infty flux}\\
    \left\langle {\cal F}^{\cal H}_{\xi}\right\rangle &= -\frac{1}{32\pi}\left\langle \int {\cal L}_k h^{AB}{\cal L}_\xi h_{AB}\, dS_0\right\rangle.\label{eq:horizon flux}
\end{align}
Here the first integral is over a surface of fixed retarded time $u$ and large radius $r$, the second is over the primary's horizon at fixed advanced time $v$, and $dS_0$ is the background surface element. In retarded polar coordinates $(u,r,\theta^A)$, ${\cal L}_t$ reduces to $\partial_u$ in the first integral; in advanced, horizon-regular coordinates $(v,r,\theta^A)$, the derivative along the horizon's generators, ${\cal L}_k$, reduces to $(\partial_v+\Omega_H\partial_\psi)$, where $\Omega_H$ is the horizon's rotation frequency and $\psi$ is the horizon-regular azimuthal coordinate. In both cases, indices are raised with the inverse unit two-sphere metric $\Omega^{AB}={\rm diag}(1,\csc^2\theta)$. 

Equations~\eqref{eq:infty flux} and \eqref{eq:horizon flux} provide a clear, intuitive form of the flux contributions ${\cal F}_\xi^{(\env\times\rm p)}$: they correspond to cross terms in the standard flux formulas, comprising products of derivatives of the two waveforms $h^{(\rm p)}_{AB}$ and $h^{(\rm env)}_{AB}$.

Equation~\eqref{eq:AveragedThirdLaw} is the central result of the leading order orbit-averaged calculation: the average rate at which work is applied to the orbital energy (or torque to the orbital angular momentum) by the environment is equal and
opposite to the average power (or torque) applied to the environment by the orbiting particle, up to the gravitational-wave flux emitted at $\mathcal{O}(\epsilon\lambda)$. In scenarios where
this flux is subdominant, such as at large separations, one can then expect to recover Newton's Third Law in averaged form,
\begin{equation}
\left\langle \frac{dP^{(\rm p)}_\xi}{dt}\right\rangle
  \approx -\left\langle F^{(\rm env)}_{\xi}\right\rangle.
\end{equation}

\section{Multiscale orbital evolution using the Third Law}
\label{sec:phasespace}
So far we have expressed our results at leading order as the slow variation of the quasi-conserved energy and angular momentum of an EMRI's orbit as it interacts with an environment along a foliation of spacelike hypersurfaces. Some additional steps are needed to put these results in the multiscale form used for modern EMRI calculations~\cite{Hinderer:2008dm,Pound:2021qin,Mathews:2025nyb,Lewis:2025ydo} and for rapid EMRI waveform models~\cite{Chua:2020stf,Katz:2021yft,Speri:2023jte,Chapman-Bird:2025xtd,Honet:2025gge}. The essential idea in this formulation is to (i) express the metric, and therefore the waveform, as a function on the orbital phase space (represented by the particle's position and momentum, for example), and (ii) choose phase-space coordinates that eliminate oscillations from the dynamics.

In this section we re-express our equations more explicitly on the orbital phase space. Our motivation is twofold: first, to demonstrate a means of evolving the orbital parameters in the presence of complex dynamical matter simulations, in a fast-to-evaluate form that isolates the key interactions between the EMRI and the environment; and second, to extend the multiscale construction with the Third Law to higher orders for conservative environments. This second focus is necessary for applying our results to the growing number of conservative-leading matter-response calculations now present in the literature. With this construction we hope to provide a formulation ready for building fast-to-evaluate EMRI-environment evolution schemes for waveform generation. We first set up the general two-parameter multiscale system and define the two classes of environments we consider.

\subsection{Phase-space formulation and environment classes}
\label{sec:multiscale_setup}

We describe a generic, eccentric, possibly inclined orbit through the orbital elements, $P_i = m_\mathrm{p}\{-\mathcal{E}, \mathcal{L}_z,{\cal Q}\}$, being the orbital energy, angular momentum, and Carter constant, respectively, and phases
$\varphi^i = \{\varphi^r, \varphi^\theta, \varphi^\phi\}$, which are the angles conjugate to the radial action, polar action, and azimuthal angular momentum \cite{Pound:2021qin, Lynch:2023gpu,Lewis:2025ydo}. Alongside these we track the slowly evolving background quantities: for the Kerr geometry the collection $M_{A} = \{ M, S\}$, with $M$ the mass of the central black hole and
$S$ its dimensionful spin; see~Eq.~(4) of \cite{Lewis:2025ydo}. We denote the first two orbital elements by $P_A = m_{\rm p}\{-{\cal E},{\cal L}_z\}$ and emphasise that these are the conserved charges $P^{(\rm p)}_t$ and $P^{(\rm p)}_\phi$ of previous sections, meaning they differ by ${\cal O}(\epsilon,\lambda)$ from the background-geodesic momenta $m_{\rm p}u_t$ and $m_{\rm p}u_\phi$; we comment further on the interpretation of the perturbed $P_i$ in Sec.~\ref{sec:boostrap} below. 

Throughout this section we adopt a post-adiabatic (PA) order counting for the forcing functions that drive the evolution of $P_i$ and $M_A$. These forcing functions have the form $\epsilon^{n+1}\lambda^m\, g^{(n,m)}_i$ and $\epsilon^{n+1}\lambda^m\, F^{(n,m)}_A$, where the extra power of $\epsilon$ reflects the slow evolution of these quantities; such a term contributes at $(n+m)$PA order in the dynamics. The corrections to the orbital frequencies, denoted $\omega^{(n,m)}_i$, carry no such shift and multiply $\epsilon^{n}\lambda^{m}$ directly. Remainders $\mathcal{O}(N)$ collect all terms that first contribute at $N$PA order or beyond. Through 1PA order, the general two-parameter system of evolution equations then reads
\begingroup\allowdisplaybreaks%
\begin{subequations}\label{eq:generalmultiscale}
\begin{align}
\frac{d\varphi^i}{dt} &= \Omega^i_{(0,0)}(P_j;M_A)\notag\\*
 &\quad + \epsilon\, \omega^i_{(1,0)}(\varphi^j, P_j;  M_A)\notag\\*
&\quad + \lambda\, \omega^i_{(0,1)}(t,\varphi^j, P_j;  M_{A})
 + \mathcal{O}(2),\\
\frac{dP_i}{dt} &= \epsilon\Bigl[\epsilon\, g_i^{(0,0)}(\varphi^j, P_j;M_A)\notag\\*
 &\quad + \lambda\, g_i^{(-1,1)}(t, \varphi^j, P_j;M_A)\notag\\*
&\quad + \epsilon^2\, g_i^{(1,0)}(\varphi^j, P_j;  M_A)\notag\\*
&\quad + \epsilon\lambda\, g_i^{(0,1)}(t,\varphi^j, P_j;  M_{A})
 + \mathcal{O}(2)\Bigr],\\
\frac{d M_{A}}{dt} &=  
\lambda\, F_{A}^{(-1,1)}(t,\varphi^j,P_j;M_A)  \notag\\*
&\quad + 
\epsilon^2\, F_{A}^{(1,0)}(\varphi^j,P_j;M_A)+ \mathcal{O}(2).
\end{align}
\end{subequations}\endgroup
These equations are easily derived following Sec.~IIB of Ref.~\cite{Lewis:2025ydo}, for example, noting the overall factor of $\epsilon$ in $dP_i/dt$ arises from the fact that our $P_i$ variables include an overall factor of $m_p$.

Before proceeding, we briefly describe the physical mechanisms that can give rise to each type of frequency or forcing function in Eq.~\eqref{eq:generalmultiscale}, beginning with the phase equation. $\Omega^i_{(0,0)}$ are the background Kerr geodesic frequencies, $\omega^i_{(1,0)}$ the corrections to the orbital frequencies due to the self-gravity of the object ($h^{(1,0)}_{\alpha\beta}$, which generates the conservative first-order self-force), and $\omega^i_{(0,1)}$ the corrections to the orbital frequencies arising from the gravitational field of the leading-order matter distribution ($h^{(0,1)}_{\alpha\beta}$ in the linear-response case). For the orbital momenta, we have the leading-order gravitational-wave flux terms arising from $g^{(0,0)}_i$ (corresponding to the dissipative self-force generated by $h^{(1,0)}_{\alpha\beta}$). Then we have the dissipative effect of the environment's leading-order gravitational field, $g^{(-1,1)}_i$; this can enter simply due to the environment's own intrinsic dynamics, such as turbulent eddies or large-scale transient spiral-arm structure, or if the matter field responds nonlinearly to the secondary's potential. $g^{(1,0)}_i$ is then the correction arising from the full vacuum first-order self-force and dissipative second-order self-force. Finally, $g^{(0,1)}_i$ is the effect on the EMRI due to the mixed-order gravitational perturbation $h^{(1,1)}_{\alpha\beta}$, which includes effects such as dynamical friction and Lindblad resonant interactions. $F_A^{(1,0)}$ accounts for the accretion of energy and angular momentum onto the central black hole by gravitational waves emitted from the secondary, and $F_A^{(-1,1)}$ for the accretion of matter onto the primary, which could potentially be significant in super-Eddington accretion flows. 

Importantly in moving to phase space we will replace our time derivative with derivatives on phase space.  A total time derivative of a function $f(t,\varphi^j,P_j;M_A)$ then becomes
\begin{align}\label{eq:phasespacechainrule}
    \mathcal{L}_t &= \partial_t + \dot\varphi^i \partial_{\varphi^i}
    + \dot P_i \partial_{P_i}
    + \dot{ M}_A \partial_{M_A},\\ \notag
    &= \partial_t +\left[\Omega^{i}_{(0,0)} + \epsilon \omega^{i}_{(1,0)} + \lambda \omega^{i}_{(0,1)} \right]\partial_{\varphi^i} \\ \notag
    &\quad + \epsilon\left[\epsilon g_{i}^{(0,0)} + \lambda g_{i}^{(-1,1)}+\epsilon^2 g_{i}^{(1,0)}+\epsilon\lambda g_{i}^{(0,1)}\right]\partial_{P_i} \\ \notag
    &\quad + \left[\lambda F_{A}^{(-1,1)}+\epsilon^2 F_{A}^{(1,0)} \right]\partial_{M_A},
\end{align}
where we have denoted $\dot{x} = dx/dt$.

For the orbital elements $P_\xi$, the direct environmental forcing function, $g_\xi^{(-1,1)}$, follows immediately from the previous sections and takes the form
\begin{equation}
    g_\xi^{(-1,1)} = \frac{1}{2}\int_{\Sigma_t} T^{\alpha\beta}_{(\rm p)}\, \mathcal{L}_{\xi} h_{\alpha\beta}^{(\env)}\, d\Sigma_0.
\end{equation}
Depending on the configuration of the matter fields, this can have a generic temporal
structure. We now define the distinction between the two classes of environments we mentioned earlier:
\begin{itemize}
\item {\bf Class-D} environments are ones that exhibit dynamical features at leading order, so that $g^{(-1,1)}_i$ is non-vanishing with generic time dependence; the environment then
perturbs the orbital elements already at 0PA order.
\item {\bf Class-C} environments are those for which the forcing function $g^{(-1,1)}_i$ vanishes identically. Notably this occurs for stationary and axisymmetric configurations, so that the environment generically affects the dynamics first at the sub-leading 1PA order, through $\omega^{(0,1)}_i$ and $g^{(0,1)}_i$.
\end{itemize}

Class-D environments correspond to the types of environments arising from dynamical simulations such as turbulent discs, which have complex temporal structure and can move in and out of coherence with the EMRI's orbital frequencies throughout the inspiral. Class-C environments generally represent EMRI-environment calculations in systems where the leading-order structure of the unperturbed matter field is stationary and axisymmetric, such as in the case of recent boson cloud and disc calculations, which give rise to fully coherent wakes or Lindblad-resonance--type structures and are of widespread interest in the field at present~\cite{Polcar:2025yto,Brito:2023pyl,Dyson:2025env,Li:2025ffh,Dyson:2026ddd,Duque:2025,Hegade:2025a,Xu:2026cky,Keijzer:2026vul,Dittmann:2026rtj,Li:2026foy}.

\subsection{Class D: dynamical environments at 0PA order}
\label{sec:BalanceTypeD}
We first examine Class-D environments. We will formalise the dynamics in this section such that, even given a complex matter interaction, our forcing functions and dynamics will be constructed so as to allow a clear online-offline split, in which expensive calculations can be performed offline and the results stored for use in fast waveform generation; this is the essential aspect that enables modern, fast EMRI waveform models. The construction will be made such that the averaged forcing functions only vary over timescales much longer than the orbital timescale but much shorter than the total radiation-reaction timescale, giving a suitable construction for fast waveform generation. The offline step in this calculation requires running a simulation at each point in the slowly varying $(P_i,M_A)$ parameter space, for a duration sufficient to capture any significant nonlinear adjustments to the environment, to reach a quasi-steady state, and to cover a significant portion of the orbital phase space. This is the means by which one must construct $T^{\mu\nu}_{(\env)}(t,P_i,\varphi^i,M_A)$ as input, noting that the environment will also in general have its own independent time dependence.

Working at $\mathcal{O}(\epsilon\lambda)$, the leading environmental order, we
truncate Eqs.~\eqref{eq:generalmultiscale} at 0PA. For simplicity, we assume $\epsilon \gg \lambda$, so that the
dynamics remains radiation-reaction dominated.
In this regime of Class-D environments,
we have
\begin{subequations}\label{eq:evoltuionequations}
\begin{align}
\frac{d\varphi^i}{dt} &= \Omega^i_{(0,0)}( P_j) + \mathcal{O}(1), \label{eq:evoltuionequations_a}\\
\frac{dP_i}{dt} &= \epsilon\Bigl[\epsilon\, g_i^{(0,0)}(\varphi^j, P_j) \notag \\
&\qquad+ \lambda\, g_i^{(-1,1)}(t, \varphi^j, P_j) + \mathcal{O}(1)\Bigr], \label{eq:evoltuionequations_b} \\
\frac{dM_A}{dt} &= \lambda F_{A}^{(-1,1)}(t, \varphi^j, P_j) + \mathcal{O}(1),  \label{eq:evoltuionequations_c}
\end{align}
\end{subequations}
 where, for simplicity, we suppress the dependence on the background parameters $M_A$.
Our next step is to transform the phase-space variables to eliminate oscillations and rapid variations on the right-hand side of these equations. At leading order this is achieved by transforming to averaged variables,
\begin{equation}
    \mathring P_i = \langle P_i \rangle, \qquad \mathring M_A = \langle M_A \rangle,
\end{equation}
where the average is performed around the time $t$,
\begin{equation}\label{eq:average}
    \left\langle\,\cdot\,\right\rangle = \lim_{T\to\infty}\frac{1}{2 T}\int_{t- T}^{t + T} \cdot\;\, dt',
\end{equation}
as in Eq.~\eqref{avg Qdot}. As above, we continue to suppress the dependence on $M_A$, which after averaging is understood to be a dependence on $\mathring M_A$. The averaged variables $\mathring P_i$ are related to the original variables by a near-identity averaging transformation of the form
\begin{multline}\label{eq:transformation to Pavg}
P_i(t,{\varphi}^j, \mathring P_j) = \mathring P_i +
\epsilon\,\Delta P^{(0,0)}_i({\varphi}^j, \mathring P_j) \\
+\lambda\,\Delta P^{(-1,1)}_i(t,{\varphi}^j, \mathring P_j) +\mathcal{O}(1),   
\end{multline}
with an analogous transformation $M_A = \mathring M_A + \lambda\,\Delta M^{(-1,1)}_A(t,\varphi^j,\mathring P_j) + \mathcal{O}(1)$ for the primary's parameters. One can easily find the transformations $\Delta P_i^{(n,m)}$ and $\Delta M_A^{(n,m)}$ by substituting Eq.~\eqref{eq:transformation to Pavg} into Eq.~\eqref{eq:evoltuionequations}, applying the chain rule~\eqref{eq:phasespacechainrule}, and choosing them to eliminate non-averaged parts of the forcing functions $g^{(n,m)}_i$ and $F^{(n,m)}_A$. From this we obtain the modified system of equations
\begin{subequations}\label{eq:phaseaverageevoltuionequations}
\begin{align}
\frac{d\varphi^i}{dt} &= \Omega^i_{(0,0)}(\mathring P_j) + \mathcal{O}(1), \label{eq:phaseaverageevoltuionequations_a}\\
\frac{d\mathring P_i}{dt} &= \epsilon\Bigl[\epsilon\, G_i^{(0,0)}(\mathring P_j) \notag \\
&\qquad + \lambda\, G_i^{(-1,1)}(t, \varphi^j, \mathring P_j) + \mathcal{O}(1)\Bigr], \label{eq:phaseaverageevoltuionequations_b}\\
\frac{d \mathring{M}_A}{dt} &= \lambda\, H_A^{(-1,1)}(t,\mathring P_j) + \mathcal{O}(1), \label{eq:phaseaverageevoltuionequations_c}
\end{align}
\end{subequations}
with the averaged forcing functions
\begin{equation}
    G^{(n,m)}_i = \left\langle g^{(n,m)}_i\right\rangle, \qquad H^{(n,m)}_A = \left\langle F^{(n,m)}_A\right\rangle.
\end{equation}
Physically, $H^{(-1,1)}_A$ is simply the window-averaged flux of the environment's energy and angular momentum through the horizon,
\begin{equation}\label{eq:H(-1,1)}
    H^{(-1,1)}_A = \Bigl\langle\oint_{\mathcal{H}_t} T^{(\env)}_{\alpha\beta}\,\xi^\alpha_{(A)}\,k^\beta\,d\mathcal{S}\Bigr\rangle,
\end{equation}
where $\mathcal{H}_t$ is the intersection of $\Sigma_t$ with the horizon (taking $\Sigma_t$ to now be horizon-penetrating), $k^\alpha = \xi^\alpha_{(t)} + \Omega_H\,\xi^\alpha_{(\phi)}$ is the horizon generator, and $d\mathcal{S}$ is the surface element.
Computing the averaged forcing functions requires a careful treatment of the averaging operation $\langle\, \cdot\,\rangle$. We take the averaging window $[t-T,\, t+T]$
to have a width
\begin{equation}
    M\ll T\ll M/\epsilon,
\end{equation}
much larger than the orbital timescale but much smaller than the
radiation-reaction timescale. Taking the $T\to\infty$ limit then corresponds to neglecting terms that are subleading in $\epsilon$.  Within each window, the orbital variables can be approximated by
\begin{align}
    \mathring{P}_i(t') &= \mathring{P}_i(t) + {\cal O}(\epsilon T,\lambda T),\label{eq:P in window}\\
    \mathring{M}_A(t') &= \mathring{M}_A(t) + {\cal O}(\lambda T),\label{eq:M in window}\\
    \varphi^i(t') &= \varphi^i(t) + \Omega^i_{(0,0)}(\mathring P_j)(t'-t) + {\cal O}(\epsilon T^2,\lambda T^2),\label{eq:phi in window}
\end{align}
with error terms that are small so long as $T=T_0/\epsilon^p$ for $0<p<1/2$, given our assumption $\epsilon\gg \lambda$. 

For the vacuum forcing function $g^{(0,0)}_i(\varphi^j,\mathring P_j)$, which has no explicit time dependence, Eqs.~\eqref{eq:P in window} and \eqref{eq:phi in window} imply that the infinite time average is equivalent to an average over the orbital phases, keeping $\mathring P_j$ fixed:
\begin{equation}
    G^{(0,0)}_i(\mathring P_j) = \frac{1}{(2\pi)^3}\oint g^{(0,0)}_i(\varphi^j,\mathring P_j) d^3\varphi.
\end{equation}
The exception to this is at resonances, where two of the frequencies $\Omega^i_{(0,0)}$ have a rational ratio~\cite{Flanagan:2010cd,Isoyama:2018sib}. In either case, such averages are a standard feature of self-force calculations, and the near-identity averaging transformations required to remove oscillatory terms are well developed~\cite{vandeMeent:2013sza,VanDeMeent:2018cgn}.

Averaging the environmental forcing functions $g^{(-1,1)}_i(t,\varphi^j,\mathring P_j)$ is more subtle due to the explicit time dependence. We evaluate these averages following Ref.~\cite{Zwick:2026thw}. For concreteness we restrict our attention to the evolution of the conserved charges $P_A=P_{\xi^A}$, which are governed by the Third Law. 

We could employ the Third Law in the form~\eqref{eq:AveragedThirdLaw}, but that form would  require solving for the metric perturbation sourced by the environment in order to evaluate the emitted fluxes. Instead, defining $G^{(n,m)}_{\xi}=\left\langle g^{(n,m)}_{\xi}\right\rangle$, we exploit the Third Law in the form~\eqref{eq:averaged 3rd law v1}, which implies
\begin{multline}\label{eq:G(-1,1)}
  G_\xi^{(-1,1)}
   \\= -\lim_{T\to\infty}\frac{1}{4T}
      \int_{t-T}^{t+T}dt'\int_{\Sigma_{t'}} T^{\alpha'\beta'}_{(\rm env)}\,{\cal L}_{\xi'}h^{(\rm p,\,Adv )}_{\alpha'\beta'}\,d\Sigma'_0.
\end{multline}
All remaining mentions of $h^{(\mathrm{p})}_{\mu\nu}$ throughout the rest of this section will be in relation to the advanced field, $h^{(\rm p, Adv)}_{\mu\nu}$. 

We evaluate the integral over $t'$ by appealing to Eqs.~\eqref{eq:P in window} and \eqref{eq:phi in window}, which allow us to hold the orbital elements $\mathring{ P}_i$ fixed within the averaging window and make the $t'$ dependence explicit in $\varphi^i$. To draw out that $t'$ dependence, we expand the secondary's metric perturbation in a discrete Fourier series in the orbital phases,
\begin{equation}
    h^{(\rm p)}_{\alpha\beta} = \sum_{\bf k}h^{(\rm p,{\bf k})}_{\alpha\beta}(\mathring P_j)\,e^{-i k_j\varphi^j},
\end{equation}
where for readability we will now also drop the explicit dependence on $\mathring{ P}_j$ in our equations. Expressing the time dependence of the environment's stress-energy tensor in terms of a short-time Fourier transform, we then obtain
\begin{equation}\label{eq:ISTFT}
     T_{(\env)}^{\alpha\beta}(t') = \frac{1}{2 \pi w(t'-t)}\int_{-\infty}^\infty \tilde T_{(\env)}^{\alpha\beta}(t,\sigma)\,e^{i \sigma t' } d\sigma ,
\end{equation}
where we have defined the short-time Fourier transform $\tilde T_{(\env)}^{\alpha\beta}(t,\sigma)$ over the local window,
\begin{equation}\label{eq:STFT_environment_stress_energy}
   \tilde T_{(\env)}^{\alpha\beta}(t,\sigma) = \int_{-\infty}^{\infty} w(t'-t) T_{(\env)}^{\alpha\beta}(t') e^{- i \sigma t'}dt'.
\end{equation}
Here $w(t'-t)$ is a window function centred on $t$, which we take to vanish outside $t'\in(t-T,t+T)$. For simplicity, we assume $w$ is a top hat centred on 0, such that $\tilde T^{(\env)}_{\alpha\beta}(t,\sigma) = \int_{t-T}^{t+T} T^{(\env)}_{\alpha\beta}(t') e^{- i \sigma t'}dt'$. The window function $w$ is then set to unity in Eq.~\eqref{eq:ISTFT}, given that we only require $T^{\alpha\beta}_{(\env)}$ within the averaging window.  

We can now proceed to explicitly evaluate the forcing function~\eqref{eq:G(-1,1)} obtained from the Third Law. Defining the coefficients
\begin{equation}
    \{\mathfrak{C}^{\bf k}_{t},\mathfrak{C}^{\bf k}_{\phi}\} := \{ -  i k_j \Omega^j_{(0,0)},  i k_{\phi} \}
\end{equation}
and substituting our Fourier expansions of the constituent fields, we obtain
\begin{align}
     G_\xi^{(-1,1)}
     &=  -\frac{1}{8\pi T}\sum_{\bf k}\mathfrak{C}^{\bf k}_\xi
     \int_{t-T}^{t+T}dt'
     \int d\sigma  \int_{\Sigma_{t'}}\tilde T^{\alpha'\beta'}_{(\rm env)} (t,\sigma)
     \notag\\
      &\qquad\times\,
      {h}^{(\rm p,{\bf k})}_{\alpha'\beta'}\,
      e^{i \sigma t' - i k_j \varphi^j(t')}\,d\Sigma'_0.
      \label{eq:pidot2}
\end{align}
Expanding the phase around the window centre as in Eq.~\eqref{eq:phi in window} 
and directly integrating over $t'$ gives
\begin{align}
      G_\xi^{(-1,1)}
    &=  -\frac{1}{4\pi}\sum_{\bf k}\mathfrak{C}^{\bf k}_\xi
     \int d\sigma  \int_{\Sigma_t} \tilde T^{\alpha'\beta'}_{(\rm env)} (t,\sigma)
      {h}^{(\rm p,{\bf k})}_{\alpha'\beta'} \notag\\
      & \times\text{sinc}[ T(\sigma - k_j\Omega^{j}_{(0,0)})  ]\,e^{i\sigma t - i k_j\varphi^j(t)}
     \,d\Sigma'_0.
      \label{eq:pidot3}
\end{align}
As in Ref.~\cite{Zwick:2026thw}, we then obtain a delta approximation for the ${\rm sinc}$  function in the long-window limit, defined in the sense of distributions:\footnote{More precisely, $\int_{-\infty}^\infty \text{sinc}(\sigma T)\psi(\sigma)d\sigma = \frac{\pi}{T}\psi(0) + {\cal O}(1/T^\infty)$ for any compactly supported test function $\psi$.}
\begin{equation}
    \text{sinc}[T(\sigma - k_j\Omega^j_{(0,0)})]\underset{T\rightarrow\infty}{\longrightarrow}\frac{ \pi}{T}\delta(\sigma - k_j\Omega^j_{(0,0)}).
\end{equation}
Finally, integrating over $\sigma$, we obtain
\begin{align}\label{eq:averagedequations}
     G_\xi^{(-1,1)}
    &=  -\frac{1}{4 T}\sum_{\bf k}\mathfrak{C}^{\bf k}_\xi
     \int_{\Sigma_t} \tilde T^{\alpha\beta}_{(\rm env)} (t,k_j\Omega^j_{(0,0)},x^i)
     \notag\\
     &\times {h}^{(\rm p,{\bf k})}_{\alpha\beta}(x^i)\,e^{-i k_j[\varphi^j(t) - \Omega^j_{(0,0)}t]}
     \,d\Sigma_0,
\end{align}
which now evaluates the local spectral decomposition of the environment's stress-energy tensor on multiples of the orbital frequencies. From the same spectral decomposition, we readily find the explicit form of the horizon flux~\eqref{eq:H(-1,1)}. Since no metric perturbation multiplies $T^{(\env)}_{\alpha\beta}$, the window average simply selects its zero-frequency component,
\begin{equation}\label{eq:H spectral}
    H^{(-1,1)}_A = \frac{1}{2T}\oint_{\mathcal{H}_t}\tilde T^{(\env)}_{\alpha\beta}(t,0,x^i)\,\xi^\alpha_{(A)}k^\beta\,d\mathcal{S}.
\end{equation}
For $G^{(-1,1)}_\xi$, we are now left with a forcing function whose main contribution arises through the term
\begin{align}
\mathfrak{T}_{\bf k}(t,\mathring{ P}_i):=\int_{\Sigma_t}
\tilde T^{\alpha\beta}_{(\env)} (t,k_j\Omega^j_{(0,0)},x^i)\,
{h}^{(\rm p,{\bf k})}_{\alpha\beta}(x^i)d\Sigma_0.
\end{align}
For a given mode number ${\bf k}$, we may consider a curve through the 
$\mathring P_i$ space, parameterised by $t$. The short time Fourier transform of $T^{\alpha\beta}_{(\env)}$ is then evaluated at the orbital harmonics $\sigma=k_j\Omega^j_{(0,0)}$, giving
$\mathfrak{T}_{\bf k}(t,\mathring{ P}_i)$ along the curve.
This provides us with our clear online-offline split, from which one can perform an amplitude-phase splitting in $t$, by simply defining
\begin{align}\label{eq:Tk frak}
\mathfrak{T}_{\bf k}(t,\mathring P_i)=\mathfrak{A}_{\bf k}(t,\mathring P_i)\,
e^{i \varphi^{\bf k}_{\env}(t,\mathring P_i)}.
\end{align}
Due to the complex-valued nature of $\mathfrak{T}_{\bf k}$, we immediately have
\begin{subequations}
\begin{align}
\mathfrak{A}_{\bf k}(t,\mathring P_i) &= \big|\mathfrak{T}_{\bf k}(t,\mathring P_i)\big|,\\
\varphi^{\bf k}_{\env}(t,\mathring P_i) &= \arg\, \mathfrak{T}_{\bf k}(t,\mathring P_i),
\end{align}
\end{subequations}
and we take $\varphi^{\bf k}_{\env}$ to be the
continuous branch obtained by unwrapping the phase along $t$, ensuring that $\varphi_{\env}^{\bf k}$ varies smoothly. Substituting these expressions
into Eq.~\eqref{eq:averagedequations} and labelling the slowly modulating orbital
phase residual as
\begin{equation}
    \delta\varphi^j_{(\rm p)}(t) = \varphi^j(t) - \Omega^j_{(0,0)}t,
\end{equation}
we obtain the full averaged dynamics,
\begin{subequations}\label{eq:averagedequationsenvdecomp}
\begin{align}
    \frac{d\varphi^i}{dt} &= \Omega^i_{(0,0)}(\mathring P_j) + \mathcal{O}(1),\\
    \frac{d \mathring{P}_i}{dt}
    &= \epsilon\biggl[\epsilon\, G_i^{(0,0)}(\mathring P_i) \notag\\
    &\quad-\frac{\lambda}{4 T}\sum_{\bf k}\mathfrak{C}^{\bf k}_{i}\mathfrak{A}_{\bf k}(t,\mathring P_j)
    e^{i\left[\varphi^{\bf k}_{\env}(t) -k_j\delta\varphi^j_{(\rm p)}(t) \right]}\notag\\
   &\quad +\mathcal{O}(1)\biggr],\label{eq:averagedequationsenvdecomp_b}\\
    \frac{d\mathring M_A}{dt} &= \lambda\, H^{(-1,1)}_A(t,\mathring P_j) + \mathcal{O}(1),
\end{align}
\end{subequations}
which can be integrated directly to obtain the secondary's evolution in a highly complex, dynamical matter field. Although we have only derived the coefficients $\mathfrak{C}^{\bf k}_{A}\mathfrak{A}_k$ for the two conserved charges $P_{\xi}$, we expect the general form \eqref{eq:averagedequationsenvdecomp_b}  to hold also for the evolution of the
averaged Carter constant.

Finally, we note that Eqs.~\eqref{eq:averagedequationsenvdecomp_b} and \eqref{eq:H spectral} appear to retain a dependence on the window time $T$, which might seem counter-intuitive given that we have taken the $T\to\infty$ limit. This dependence is, however, compensated by an implicit dependence of $\mathfrak{A}_{\bf k}$ and $\tilde T^{(\env)}_{\alpha\beta}(t,0,x^i)$ on $T$: from Eqs.~\eqref{eq:STFT_environment_stress_energy} and \eqref{eq:Tk frak}, both grow linearly with $T$ for any contribution to $T^{\alpha\beta}_{(\env)}$ that remains coherent with an orbital harmonic (or, for the horizon flux, is non-oscillatory) across the averaging window, so that their ratio remains finite.
\subsection{Class C: stationary, axisymmetric led environments at 1PA order}
\label{sec:BalanceTypeC}
Working to 1PA, we now restrict to environments that are stationary and axisymmetric at leading order.
We expect that an application of the Third Law to scenarios such as these will act as an important tool in constructing fast-to-evaluate, parameter-space-covering, precision waveform models for EMRIs in environments. In particular this construction leads to an easy-to-evaluate evolution scheme for a number of the self-consistent matter-EMRI calculations in conservative environments already present in the literature. 

Owing to the Class-C nature of the environment, there is no order-$\epsilon\lambda$ term in the evolution of $P_i$. This is immediately clear from Eq.~\eqref{eq:standardforce}: if $h^{(\env,0,1)}_{\alpha\beta}$ is stationary and axisymmetric, then ${\cal L}_\xi h^{(\env,0,1)}_{\alpha\beta}$ vanishes in the integrand. Although that argument only applies to $P_\xi$, the same holds true for a perturbed Carter constant away from resonances (by virtue of the Kolmogorov-Arnold-Moser theorem~\cite{Gair:2007kr,Lukes-Gerakopoulos:2010ipp}). Moreover, a stationary, axisymmetric solution has no net flux of matter or gravitational waves through the horizon, enforcing  $F^{(-1,1)}_A=0$ in our evolution equations. The environmental effects on the orbit then enter in the phases and evolution of the slowly varying orbital parameters at 1PA, and the multiscale system is Eqs.~\eqref{eq:generalmultiscale} with $g^{(-1,1)}_i = 0 = F^{(-1,1)}_A$ and with the order-$\lambda$ term in $d\varphi^i/dt$ easily determined from $h^{(\env,0,1)}_{\mu\nu}$. 

The aim of this section is now to apply the Third Law to determine the remaining 1PA environmental effect: the order-$\epsilon^2\lambda$ term in $dP_i/dt$. In doing so, we will represent the orbital evolution equations in an efficient, easy-to-solve manner which circumvents the need to explicitly calculate the metric contribution due to the perturbed environment, $h^{(1,1)}_{\mu\nu}$, instead only requiring the metric contribution from the matter background $h^{(\text{env},0,1)}_{\mu\nu}$. Additionally, there has been some ambiguity in recent years as to the very nature of the correct EMRI forcing function. Our results will therefore not only be easy to evaluate but also clarify its form.

Our construction is particularly motivated by recent advances in directly solving the linearised Einstein field equations in stationary, axisymmetric systems (see in particular public codes by Fernandes~\cite{Fernandes:2025vxg}), making the construction of $h^{(\env,0,1)}_{\mu\nu}$ in these contexts now a relatively straightforward task.

\subsubsection{Bootstrapping the Third Law}\label{sec:boostrap}

Our results for the Third Law in Sec.~\ref{sec:BalanceFirst} only apply to the order-$\epsilon\lambda$ term in $dP_\xi/dt$, which vanishes in the present case.  To determine the subleading, order-$\epsilon^2\lambda$ term in $dP_i/dt$, we now bootstrap our leading-order results to higher order.

We do this by considering perturbations on top of an exact stationary axisymmetric solution to the Einstein equations coupled to matter,
\begin{align}
    G_{\mu\nu}[\hat g] = 8\pi \lambda\hat T_{\mu\nu}^{(\env)},\quad\hat\nabla_{\mu}\hat T^{\mu\nu}_{(\env)} = 0.
\end{align}
Here the black-hole--matter metric can be expanded as 
\begin{equation}
    \hat g_{\mu\nu} = g_{\mu\nu} + \lambda h^{(\env,0,1)}_{\mu\nu} + {\cal O}(\lambda^2), 
\end{equation}
and, similarly,  $\lambda\hat T^{\mu\nu}_{(\env)} = \lambda T^{\mu\nu}_{(\env,0,1)} + {\cal O}(\lambda^2)$. 
For the moment we treat these fields nonperturbatively, and we denote objects defined on $\hat g_{\mu\nu}$ 
with a hat. We then perturb the coupled system by introducing the secondary, leading to linearized Einstein equations of the form
\begin{align}\label{eq:linear_einstein_boostrap}
\hat {\cal E}_{\mu\nu}{}^{\alpha\beta}\hat{\bar  h}_{\alpha\beta}^{(\rm p)}  = 8 &\pi \hat T_{\mu\nu}^{(\rm p)},\\
\hat {\cal E}_{\mu\nu}{}^{\alpha\beta} \delta \bar{h}_{\alpha\beta}^{(\rm env)}  = 8 &\pi \delta T_{\mu\nu}^{(\env)},
\end{align}
where $\epsilon \hat T^{(\mathrm{p})}_{\mu\nu}=\epsilon T^{(\mathrm{p},1,0)}_{\mu\nu}+\epsilon\lambda T^{(\mathrm{p},1,1)}_{\mu\nu} + {\cal O}(\epsilon\lambda^2)$ is the point-particle stress-energy tensor specified on the exact stationary axisymmetric black-hole matter solution $\hat g_{\alpha\beta}$; cf. Eq.~\eqref{eq:Tp11}. The second source, $\delta T^{(\env)}_{\alpha\beta}$, is the perturbation to the matter field by the secondary potential, which scales as $\epsilon\lambda$. For simplicity, we assume $\delta T^{(\env)}_{\alpha\beta}$ is bounded at the particle; we highlight the importance of this assumption below.

Because the environment effectively enters at zeroth order in this construction, one must also simultaneously solve the coupled matter equations fixed by
\begin{align}
\bm\nabla_{\mu}  \left(\epsilon\delta T^{\mu\nu}_{(\env)}+ \hat T^{\mu\nu}_{(\env)} \right)=0,\label{eq:div(dT+That)}\\
\tilde\nabla_{\mu} T^{\mu\nu}_{(\mathrm{p})} =0,
\end{align}
where the full metric is given by
\begin{equation}
   \bm g_{\mu\nu}:= \hat g_{\mu\nu}+   \epsilon\hat h^{(\rm p)}_{\mu\nu}+\epsilon\lambda\delta h^{(\env)}_{\mu\nu},
\end{equation}
and $T^{\mu\nu}_{(\mathrm{p})}$ is the full particle stress-energy from Eq.~\eqref{eq:TDet}. 
As before, $T^{\mu\nu}_{(\mathrm{p})}$ is conserved with respect to the effective regular metric
\begin{equation}
   \tilde g_{\mu\nu}:= \hat g_{\mu\nu}+ \hat h^{(\rm R,p)}_{\mu\nu}+ \delta h^{(\env)}_{\mu\nu},
\end{equation}
in which the singular part of the particle's field has been removed, $h^{(\rm R,p)}_{\mu\nu}:= h^{(\rm p)}_{\mu\nu} - h^{(\rm S,p)}_{\mu\nu}$. This is the first instance where the boundedness of $\delta T^{\mu\nu}_{(\env)}$ comes into play: it ensures that $\delta h^{(\env)}_{\mu\nu}$ is $C^1$ at the particle, meaning it directly enters the particle's stress-energy and equation of motion without requiring regularisation. We also note that this boundedness implies the environment's conservation equation~\eqref{eq:div(dT+That)} remains integrable at order $\epsilon^2$.

This setup is similar to that of the first self-consistent calculations of the EMRI--matter
response~\cite{Cardoso:2022whc,Duque:2023seg}. However, both the exact black-hole--matter solution and the
regularised particle field $h^{(\rm R,p)}_{\mu\nu}$ will serve here only as derivational tools and do not appear in our final
expressions. Before deriving our bootstrapped Third Law in this setting, we first highlight the freedom in defining $P^{(\mathrm{p})}_\xi$ beyond leading order. For example, we could define the momenta from the ``black hole plus environment'' metric $g^{(\env)}_{\mu\nu}=\hat g_{\mu\nu} + \delta h^{(\env)}_{\mu\nu}$, together with a ``particle in environment'' stress-energy $\epsilon T^{\mu\nu}_{(\text{p-env})}$, defined analogously to Eq.~\eqref{eq:TDet} with $\tilde g_{\mu\nu}$ replaced by $g^{(\env)}_{\mu\nu}$. In that case, the momenta would be $P^{(\text{p-env})}_\xi =\int_{\Sigma_t} T^{\mu\nu}_{(\text{p-env})} g^{(\env)}_{\nu\rho}\xi^\rho d\Sigma^{(\env)}_{\mu}$. Alternatively, we can include the secondary's self-field by defining the momenta from the effective metric $\tilde g_{\mu\nu}$ and the full stress-energy $T^{\mu\nu}_{(\rm p)}$, as in 
\begin{equation}
\tilde P^{(\mathrm{p})}_\xi \coloneqq\int_{\Sigma_t} T^{\mu\nu}_{(\mathrm{p})} \tilde g_{\nu\rho}\xi^\rho d\tilde \Sigma_{\mu}. 
\end{equation}
Either definition provides suitable variables to describe the orbital evolution, but the different momenta satisfy different evolution equations. On the other hand, defining the momenta in terms of the full physical metric $\bm g_{\mu\nu}$, as in Eq.~\eqref{Q def}, would lead to divergent integrals at the orders of interest in this section.

For concreteness, we work with the momenta in the effective metric, $\tilde P^{(\mathrm{p})}_\xi$. We also introduce the power (or torque) functional on the background $\hat g_{\alpha\beta}$,
\begin{equation}\label{eq:work_functional_hat}
    \hat{\mathcal{W}}_{\xi}[T,h]:= \lim_{T\to\infty}\frac{1}{4T}\int_{V_T}
     T^{\alpha\beta}\,{\cal L}_{\xi}h_{\alpha\beta}\, d\hat V.
\end{equation}  
This is in contrast to $\mathcal{W}_{\xi}[T,h]$, which will remain defined with respect to the original Kerr metric as in Eq.~\eqref{eq:work_functional}.

Since the particle's stress-energy is conserved with respect to the effective metric, and since ${\cal L}_\xi\hat g_{\mu\nu}=0$, our initial derivation of the Third Law applies directly, replacing $\bm g_{\mu\nu}$ with $\tilde g_{\mu\nu}$ and $T^{\mu\nu}$ with $T^{\mu\nu}_{(p)}$ in the derivation leading to Eq.~\eqref{eq:fullQdot}. We then arrive at the functional form
\begin{equation}\label{eq:bootstrap_roc_charge}
\left<\frac{d\tilde{P}^{(\rm p)}_\xi}{dt} \right>= \lim_{T\rightarrow\infty}\frac{\epsilon^2}{4 T}\int_{V_T} \hat T^{\alpha\beta}_{(\rm p)}{\cal L}_\xi\!\left(\hat h^{(\rm R,p)}_{\alpha\beta} + \lambda\delta h^{(\env)}_{\alpha\beta}\right)d\hat V
\end{equation}
for the averaged power or torque on the secondary at leading nontrivial order. Here $\hat g_{\mu\nu}$ plays the role of $g_{\mu\nu}$ from the earlier derivation, and we have discarded the contributions of $h^{(\rm R,p)}_{\alpha\beta}$ to $T^{\alpha\beta}_{(\rm p)}$ and to $\tilde g_{\mu\nu}$ as they would enter Eq.~\eqref{eq:bootstrap_roc_charge} at order $\epsilon^3$. 

With the explicit form~\eqref{eq:bootstrap_roc_charge}, we can split the Detweiler-Whiting regular Green's function as
\begin{equation}
    \hat G^{(\rm R)}_{\mu\nu\alpha'\beta'}  = \hat G^{(\rm Rad)}_{\mu\nu\alpha'\beta'} + \hat G^{(\rm R,\rm Sym)}_{\mu\nu\alpha'\beta'}.
\end{equation}
The radiative component of the regularised Green's function
is then equal to the full radiative Green's function by construction~\cite{Detweiler:2002mi,Poisson:2011nh}. Following steps analogous to those in Eq.~\eqref{<Qdot>}, one finds that $\hat G^{(\rm R,Sym)}_{\mu\nu\mu'\nu'}$ does not contribute to the particle--particle terms in Eq.~\eqref{eq:bootstrap_roc_charge}, such that the expression reduces to
\begin{equation}\label{eq:bootstrap_roc_charge_rad}
\left<\frac{d\tilde{P}^{(\rm p)}_\xi}{dt} \right>= \epsilon^2\hat{\mathcal{W}}_{\xi}[\hat T_{\rm (p)}, \hat h^{(\rm p, Rad)}+ \lambda\delta h^{(\rm env)}].
\end{equation}
Consequently, our construction of the averaged Third Law only requires the construction of \emph{radiative} particle fields and does not require any explicit calculation of the Detweiler-Whiting singular-regular split~\cite{Detweiler:2002mi} in the exact $\hat g_{\alpha\beta}$ spacetime. Notably we could equally have used the softened particle as in Sec.~\ref{sec:nonlinear_response} to arrive at this point: once the problem is reduced
to integrals against the radiative Green's function, which is smooth at the particle, the limit of zero softening is well defined. Hence we would arrive at the same form as Eq.~\eqref{eq:bootstrap_roc_charge_rad}.

Applying the same steps as led to Eq.~\eqref{eq:AveragedThirdLaw}, we now readily
obtain our averaged Third Law on the black-hole--matter background:
\begin{equation}\label{eq:third_law_exact_background}
\left<\frac{d\tilde{P}^{(\rm p)}_\xi}{dt} \right>= -\epsilon^2 \Bigl[\lambda\left<\hat{F}^{(\env)}_{\xi} \right> + \lambda\left<\hat{\mathcal{F}}^{(\env\times{\rm p})}_{\xi}\right> +\left<\hat{\mathcal{F}}^{(\mathrm{p}\times \mathrm{p})}_{\xi}\right>\Bigr],
\end{equation}
where we have defined
\begin{subequations}
\begin{align}
    \bigg<\hat{F}_{\xi}^{(\env)} \bigg>&:= \hat{\mathcal{W}}_{\xi}[\delta T_{\rm (\env)}, \hat h^{(\rm p,Ret)}],\\
    \left<\hat{\mathcal{F}}_{\xi}^{(\env\times{\rm p})}\right> &:= -2\hat{\mathcal{W}}_{\xi}[\delta T_{\rm (\env)}, \hat h^{(\rm p,Rad)}],\\
    \left<\hat{\mathcal{F}}_{\xi}^{(\mathrm{p}\times \mathrm{p})}\right> &:= -\hat{\mathcal{W}}_{\xi}[ \hat T_{\rm ( p)}, \hat h^{(\rm p,Rad)}].
\end{align}
\end{subequations}
Since we started with the effective rather than full metric, all terms are constructed from integrable expressions on $\hat g_{\alpha\beta}$. 

If one were to use the form~\eqref{eq:third_law_exact_background} directly, one would need to
calculate fluxes and metric perturbations on a different, non-Kerr spacetime for each new environment considered (and for every value of the parameters describing any particular type of environment). We can put the result in a more practical form by re-expanding our expression in $\lambda$, such that all perturbations are on a Kerr background. One way of doing that would be to express $\hat{\cal F}^{(\env\times \rm p)}_\xi$ and $\hat{\cal F}^{(\mathrm{p}\times \mathrm{p})}_\xi$ as a total asymptotic flux, as in Eqs.~\eqref{eq:total flux}--\eqref{eq:horizon flux}, and then expand in powers of $\lambda$. However, the resulting flux expressions would involve the subleading metric perturbation $h^{(1,1)}_{\mu\nu}$. In the
next subsection, we reformulate our result to instead arrive at an easier-to-evaluate form of the force that requires only leading-order metric perturbations on the Kerr spacetime.

Before proceeding, we return to the choice of perturbed $P_\xi$. If we had used $P^{(\text{p-env})}_\xi$ rather than $\tilde P^{(\rm p)}_\xi$ (i.e., had we omitted the self-field), then the same derivation would apply, but without the $\hat h^{(\rm R,p)}_{\mu\nu}$ term in Eq.~\eqref{eq:bootstrap_roc_charge}. The $\epsilon^2\lambda$ term in $\left<dP^{(\text{p-env})}_\xi/dt\right>$ would then be given by just the first two terms in Eq.~\eqref{eq:third_law_exact_background}, differing from $\left<d\tilde P^{(\text{p})}_\xi/dt\right>$ by the linear-in-$\lambda$ terms that arise in the self-interaction $\left<\hat{\mathcal{F}}^{(\mathrm{p}\times \mathrm{p})}_{\xi}\right>$ once $\hat g_{\mu\nu}$ is expanded in $\lambda$.\footnote{Importantly, choosing $P^{(\text{p-env})}_\xi$ does not simplify the dynamics overall; it instead moves this dependence into the frequency corrections $\omega^i_{(0,1)}$ in the evolution equations for the phases.} Unpacking that subleading term is the main goal of the next subsection.

\subsubsection{Reduction to Kerr quantities}
Rather than rewriting the radiative-field terms as asymptotic fluxes, we now directly expand Eq.~\eqref{eq:third_law_exact_background} in $\lambda$, retaining the leading-order environmental effects at
$\mathcal{O}(\epsilon^2\lambda)$. The first two terms in Eq.~\eqref{eq:third_law_exact_background} reduce to Kerr quantities
straightforwardly, since $\delta T_{(\env)}$ is already of leading order, i.e.,
$\epsilon\delta T^{(\env)}_{\mu\nu}=\epsilon\lambda T^{(\env,1,1)}_{\mu\nu}+\mathcal{O}(\epsilon\lambda^2,\epsilon^2\lambda)$, so no new cross terms arise. Hence the first two contributions of Eq.~\eqref{eq:third_law_exact_background} take the form
\begin{align}\label{eq:1PA_Third_law_kerr_T1}
    \left<F^{(\rm env)}_{\xi(2,1)}\right> &:=  \mathcal{W}_{\xi}[T_{(\env,1,1)},h^{(\rm p,Ret, 1,0)}],\\
\label{eq:1PA_Third_law_kerr_T2}
    \left<\mathcal{F}^{(\env\times{\rm p})}_{\xi(2,1)}\right> &:= -2\mathcal{W}_{\xi}[T_{(\env,1,1)},h^{(\rm p,Rad, 1,0)}],
\end{align}
at $\mathcal{O}(\epsilon^2 \lambda)$, where we note that in this context $\epsilon h^{(\mathrm{p})}_{\mu\nu} = \epsilon h^{(\mathrm{p},1,0 )}_{\mu\nu} +\epsilon\lambda h^{(\mathrm{p},1,1 )}_{\mu\nu} + \mathcal{O}(\epsilon\lambda^2,\epsilon^2\lambda) $. The final flux term in Eq.~\eqref{eq:third_law_exact_background}, however, requires additional care. At $\mathcal{O}(\epsilon^2\lambda)$, the expansion of
$\left<\mathcal{F}^{({\rm p}\times{\rm p})}\right>$ initially takes the form
\begin{align}\label{eq:GW_flux_thrdlaw_1PA_T1}
     \left<\mathcal{F}^{({\rm p}\times{\rm p})}_{\xi(2,1)}\right>
     ={}&-\mathcal{W}_{\xi}[T_{(\rm p ,1,0)}, h^{(\rm p, Rad, 1, 1)}]\notag\\
     &-\mathcal{W}_{\xi}[T_{(\rm p ,1,1)}, h^{(\rm p, Rad, 1, 0)}]\notag \\
     &-\mathcal{W}^{(1)}_{\xi}[T_{(\rm p ,1,0)}, h^{(\rm p, Rad, 1, 0)}],
\end{align}
where we have defined the first-order variation of the power (or torque) functional~\eqref{eq:work_functional} as
\begin{align}\label{eq:work_functional_variation}
    \mathcal{W}^{(1)}_{\xi}[T,h]&:= \lim_{T\to\infty}\frac{1}{4T}\int_{V_T}
     T^{\alpha\beta}\,{\cal L}^{(1)}_{\xi}h_{\alpha\beta}\,dV_0\notag\\
     &\quad+ \lim_{T\to\infty}\frac{1}{8T}\int_{V_T}
    h^{(0,1)}\, T^{\alpha\beta}\,{\cal L}_{\xi}h_{\alpha\beta}\,dV_0.
\end{align}
Here $h^{(0,1)}:=g^{\mu\nu}h^{(\env,0,1)}_{\mu\nu}$ is the trace of the environment's leading-order metric perturbation, which arises from the expansion of the volume element, $\sqrt{-\hat g}=\sqrt{-g}\,(1+\tfrac{1}{2}\lambda h^{(0,1)})$. When applied to the point-particle perturbation, the variation of the Lie derivative is
\begin{equation}
    \mathcal{L}^{(1)}_{t}h^{(\rm p,1,0)}_{\mu\nu} =  \omega_{(0,1)}^{i}\partial_{\varphi^i}h^{(\rm p,1,0)}_{\mu\nu}, \qquad \mathcal{L}^{(1)}_{\phi}h^{(\rm p,1,0)}_{\mu\nu} = 0,
\end{equation}
which arises from the multi-scale expansion~\eqref{eq:phasespacechainrule}. 

Equation~\eqref{eq:GW_flux_thrdlaw_1PA_T1} is very nearly in an easily calculable form, involving only leading-order point-particle metric perturbations on the
Kerr metric, with the exception of the term $h^{(\rm p,Rad, 1,1)}_{\mu\nu}$. The final hurdle is to express this contribution in terms of
leading-order point-particle metric perturbations on the Kerr metric. The key subtlety is that all of our metric perturbations were initially defined with respect to the Green's function of the exact metric $\hat g_{\alpha\beta}$. Owing
to the perturbative nature of our construction, after expanding in $\lambda$, any leading-order perturbation $h^{(1,0)}_{\mu\nu}$ constructed initially on $\hat g_{\alpha\beta}$ is identical to one constructed from the Kerr Green's function with the same source.
This is not the case for the subleading perturbations $h^{(1,1)}_{\mu\nu}$. We find the correct radiative piece of $h^{(\rm p,1,1)}_{\mu\nu}$, corresponding to the linear-in-$\lambda$ term in $\hat h^{(\rm p,Rad)}_{\mu\nu}$, by appropriately expanding the field equation~\eqref{eq:linear_einstein_boostrap}.

Denoting the linearised Einstein operator on $\hat g_{\alpha\beta}$ in
Lorenz gauge by $\hat{\mathcal{E}}$, we observe that the operator's expansion in $\lambda$ reads
\begin{align}
     \hat{\mathcal{E}}[\cdot]= \mathcal{E}[\cdot]  + 2 \lambda \left(   \delta^2G[h^{(\env, 0,1)},\cdot] + \frac{1}{2}\delta^{(1)}G[\cdot]\right)+\mathcal{O}(\lambda^2),
\end{align}
where $\mathcal{E}$ is now defined on Kerr, $\delta^2G$ is the standard second-order variation of the Einstein tensor, and $\delta^{(1)}G$ is the correction to the linearised operator arising from the $\mathcal{O}(\lambda)$ shift $\omega^i_{(0,1)}$ of the orbital frequencies on phase space. Now substituting this expansion along with $\hat h^{(\rm p,Ret/Adv)}_{\mu\nu} =  h^{(\rm p,Ret/Adv,1,0)}_{\mu\nu} + \lambda h^{(\rm p,Ret/Adv,1,1)}_{\mu\nu} + {\cal O}(\lambda^2)$ into Eq.~\eqref{eq:linear_einstein_boostrap}, we can construct the \emph{retarded} and \emph{advanced} fields on Kerr in their Green's-function representations, which take the form\footnote{We observe that though $T^{\mu\nu}_{(\rm p,1,0)}$ and $T^{\mu\nu}_{(\rm p,1,1)}$ depend on the secondary's orbit, they are identical, as functions on phase space, between the retarded and advanced calculations. In the multiscale expansion, sources and solutions are interpreted as functions on the phase space $(\varphi^i,P_i,M_A)$~\cite{Mathews:2025nyb,Lewis:2025ydo}; while $T^{\mu\nu}_{(\rm p,1,n)}$ would be different functions of $t$ between the retarded and advanced cases in an ordinary 3+1 treatment of the problem, they remain the same functions on phase space. The fact that the orbital trajectories differ then manifests itself in the source term $\delta^{(1)}G_{\mu\nu}$.}
\begin{align}
    h^{(\rm p,Ret,1,1)}_{\mu\nu} &= G^{(\rm Ret)}_{\mu\nu}\big[T_{(\rm p,1,1)}-Q^{(\rm Ret)}\big],\\
    h^{(\rm p,Adv,1,1)}_{\mu\nu} &= G^{(\rm Adv)}_{\mu\nu}\big[T_{(\rm p,1,1)}-Q^{(\rm Adv)}\big].
\end{align}
Here $G^{(\rm X)}_{\mu\nu}[\cdot]:=\int G^{(\rm X)}_{\mu\nu\alpha'\beta'}(\cdot)^{\alpha'\beta'}\,dV'_0$ denotes a Kerr Green's function acting on a given source (with the factor of $8\pi$ absorbed into $G^{(\rm X)}_{\mu\nu\alpha'\beta'}$, as throughout), and for convenience we have written $Q^{(\rm X)}:=Q[h^{(\rm p,X,1,0)}]$ for ${\rm X}\in\{{\rm Ret},{\rm Adv},{\rm Sym},{\rm Rad}\}$, given explicitly by
\begin{align}
    8\pi Q^{(\rm X)} &= 2\delta^2G[h^{(\env,0,1)},h^{(\rm p,X,1,0)}]\notag\\
    &\quad+\delta^{(1)}G[h^{(\rm p,X,1,0)}].
\end{align}
We can then construct the
radiative field as $h_{\mu\nu}^{(\rm p,Rad,1,1)} = (h_{\mu\nu}^{(\rm p,Ret,1,1)}-h_{\mu\nu}^{(\rm p,Adv,1,1)})/2$. Expressing the retarded and advanced Green's functions as linear combinations of the \emph{symmetric} and \emph{radiative}
Green's functions, one readily arrives at
\begin{align}\label{eq:1PA_Third_law_kerr_T3}
     h^{(\rm p,Rad,1,1)}_{\mu\nu} ={}&
  G^{(\rm Rad)}_{\mu\nu}[T_{(\rm p,1,1)}]
  - G^{(\rm Sym)}_{\mu\nu}[Q^{(\rm Rad)}] \notag \\
  &
     - G^{(\rm Rad)}_{\mu\nu}[Q^{(\rm Sym)}].
\end{align}

Substituting Eq.~\eqref{eq:1PA_Third_law_kerr_T3} into Eq.~\eqref{eq:GW_flux_thrdlaw_1PA_T1} and again employing the same types of manipulations as in Eq.~\eqref{<Qdot>}, we arrive at a directly calculable form for the particle--particle contribution to the environmental forcing of the Kerr orbital elements,
\begin{align}\label{eq:GW_flux_thrdlaw_1PA_T3}
     \left<\mathcal{F}^{({\rm p}\times{\rm p})}_{\xi(2,1)}\right>
     ={}&\mathcal{W}_{\xi}[Q^{(\rm Sym)}, h^{(\rm p, Rad, 1, 0)}]\notag\\
     &-\mathcal{W}_{\xi}[Q^{(\rm Rad)}, h^{(\rm p, Sym, 1, 0)}]\notag\\
     &-2\mathcal{W}_{\xi}[T_{(\rm p ,1,1)}, h^{(\rm p, Rad, 1, 0)}]\notag\\
     &-\mathcal{W}^{(1)}_{\xi}[T_{(\rm p ,1,0)}, h^{(\rm p, Rad, 1, 0)}],
\end{align}
where all contributions are well defined and convergent.\footnote{The last three terms can be evaluated numerically as written. In the first, $Q[\cdot]$ acts on the singular field $h^{(\rm p,Sym,1,0)}_{\mu\nu}$ and is understood in the sense of distributions: integrating its derivatives by parts onto the smooth field $\mathcal{L}_\xi h^{(\rm p,Rad,1,0)}_{\mu\nu}$ gives an absolutely convergent integral.}

We hence arrive at the final expression for the averaged environmental forcing of the Killing charges, written in terms of Kerr quantities as
\begin{align}\label{eq:g01_final_1PA}
   \big\langle g^{(0,1)}_{\xi}\big\rangle={}&-\big[\big\langle F^{(\env)}_{\xi(2,1)}\big\rangle+\big\langle\mathcal{F}^{(\env\times{\rm p})}_{\xi(2,1)}\big\rangle +\big\langle\mathcal{F}^{(\mathrm{p}\times \mathrm{p})}_{\xi(2,1)}\big\rangle\big].
\end{align}
Note that here, all quantities, including metric perturbations and stress-energies, are treated as functions on the binary's phase space and expanded for small $\epsilon$ and small $\lambda$ at fixed values of the phase-space variables $(\varphi^i,P_i,M_A)$ (i.e., the equations are formulated in the ``fixed-constants-of-motion'' gauge~\cite{Piovano:2024yks,Mathews:2025nyb}). 
As explained at the end of the preceding subsection, the contribution~\eqref{eq:GW_flux_thrdlaw_1PA_T3} to the orbital evolution is associated with the self-field contributions to the particle's energy and angular momentum in our choice of variables. 

Through Eqs.~\eqref{eq:1PA_Third_law_kerr_T1}, \eqref{eq:1PA_Third_law_kerr_T2}, and \eqref{eq:GW_flux_thrdlaw_1PA_T3}, all environmental terms can be constructed straightforwardly from point-particle metric perturbations on the Kerr geometry, the smooth, stationary, axisymmetric metric perturbation of the environment, and self-consistently computed matter-response profiles, i.e., the fields $h^{(\rm p,1,0)}_{\mu\nu}$, $h^{(\env,0,1)}_{\mu\nu}$, and $T_{\mu\nu}^{(\env,1,1)}$.
Finally, from the structure of the near-identity transformations~\cite{VanDeMeent:2018cgn}, the oscillatory parts of the 1PA forcing functions $g^{(0,1)}_i$ and $g^{(1,0)}_i$ affect the orbital phases only at 2PA order, so the full 1PA evolution requires only their averages; the full leading-order forcing $g^{(0,0)}_i$ and the frequency corrections $\omega^i_{(1,0)}$ and $\omega^i_{(0,1)}$. The environmental frequency correction $\omega^i_{(0,1)}$ is itself an explicit function of $h^{(\env,0,1)}_{\mu\nu}$, so the full 1PA environmental dynamics can be evolved without computing any gravitational perturbation at $\mathcal{O}(\epsilon\lambda)$:
\begin{subequations}\label{eq:Kerr_evolution_final_1PA}
\begin{align}
    \frac{d\varphi^i}{dt} &= \Omega^i_{(0,0)}+\epsilon\, \omega^i_{(1,0)} + \lambda\, \omega^i_{(0,1)},\\
    \frac{dP_i}{dt} &= \epsilon^2 \Big[ g_i^{(0,0)} + \epsilon\left< g_i^{(1,0)}\right> + \lambda\left< g_i^{(0,1)}\right>\Big],\\
    \frac{d M_{A}}{dt} &= \epsilon^2\, F_{A}^{(1,0)},
\end{align}
\end{subequations}
where we have implicitly performed a small transformation to eliminate the oscillatory contributions of $g^{(1,0)}_i$ and $g^{(0,1)}_i$. 
We stress that with our current derivation $\langle g^{(0,1)}\rangle$ is only available for the charges associated with Killing symmetries, and further work is required to extend this prescription to the evolution of the Carter constant, and hence to give a complete prescription for generic orbits. For equatorial orbits in environments that are reflection symmetric about the equatorial plane, the orbit remains equatorial and the Killing charges are the only orbital elements. Equation~\eqref{eq:g01_final_1PA} then provides the full construction of the environmental forcing without the need to calculate any gravitational perturbations beyond leading order, $\mathcal{O}(\epsilon)$ or $\mathcal{O}(\lambda)$, and gives the complete environmental contribution to the evolution of the EMRI's orbital elements at 1PA order.

\subsubsection{Discussion in the context of previous EMRI-matter calculations}

Beyond circumventing the computational burden of calculating the gravitational perturbation at $\mathcal{O}(\epsilon\lambda)$, a major motivation of this work is to clarify the true form of the environmental forcing on the EMRI in curved spacetime. Several approaches, each assuming balance laws or approximating different aspects of the force, have existed in the literature for some years now, which we will briefly review. We can think of the force as arising from two distinct pieces: the matter interactions, i.e. terms involving $T^{\mu\nu}_{\rm (env,1,1)}$, and terms arising from gravitational fluxes and interactions. Throughout the relativistic EMRI-in-environments literature, the work done by the matter on the EMRI and the matter fluxes to infinity and through the horizon have been largely conflated. It is vital to note that, for a general environment, these quantities are disconnected: if the environment has any internal dissipation mechanism, be it through viscous shearing, cooling, or additional bound states that can store energy, the fluxes at infinity will, in general, not give a meaningful measure of the local interaction between the EMRI and the matter.

The first self-consistent EMRI--matter calculations~\cite{Cardoso:2022whc,Duque:2023seg} built perturbations on exact black-hole--matter spacetimes, similar to our construction in Sec.~\ref{sec:boostrap} before expanding in $\lambda$. These first works correctly obtained the gravitational flux terms in Eq.~\eqref{eq:third_law_exact_background}. Where the direct matter interactions have been included, they have generally been assumed to follow from the $\delta T^{\mu\nu}_{(\env)}$ fluxes to infinity and through the horizon~\cite{Brito:2023pyl,Dyson:2025env,Keijzer:2026vul,Xu:2026aic,Xu:2026cky,Li:2025ffh,Datta:2025ruh}. However, even for simple environments such as boson clouds this balance remains unproven. Once the expansion in $\lambda$ is performed, calculating the gravitational field becomes more involved, and most practical calculations have neglected the gravitational-wave contributions~\cite{Brito:2023pyl,Dyson:2025env,Xu:2026aic,Xu:2026cky,Li:2025ffh,Dyson:2026ddd,Dittmann:2026rtj}. Moreover, prior to this work there was no justification for why the additional gravitational-wave emission produced by the environment should enter the EMRI's dynamics, beyond the simple argument that the energy carried away must come from the orbit; since the environment introduces an entirely new degree of freedom in which energy can be deposited, this was far from guaranteed. Our calculation obtains these contributions naturally from first principles through the conservation of the point particle stress energy tensor on $\hat g_{\alpha\beta}+ h^{\rm(R,p)}_{\alpha\beta} +\delta h^{(\env)}_{\alpha\beta}$.

It was recently realised in the relativistic setting by some of us~\cite{Dyson:2026ddd}, and later clarified in~\cite{Dittmann:2026rtj}, that, as in the Newtonian planetary-migration literature, the relevant quantity for the direct interaction is the integrated matter torque rather than the fluxes; its balance with the matter fluxes was established numerically for an adiabatic perfect fluid. However, both works identified the torque exerted by the secondary on the matter rather than the torque exerted by the matter on the secondary, the na\"{\i}ve, acausal form of the Third Law discussed in Sec.~\ref{sec:instantaneous}. The correct quantity pairs the matter perturbation with the \emph{advanced}, rather than the \emph{retarded}, point-particle metric perturbation, and as a result these works missed the radiative matter-torque term in Eq.~\eqref{eq:1PA_Third_law_kerr_T2}.

Some works~\cite{Rahman:2025mip,Keijzer:2026vul,Li:2026foy} have calculated the additional gravitational-wave flux at $\mathcal{O}(\epsilon^2\lambda)$, finding that it can be up to two orders of magnitude larger than the matter flux close to merger, indicating that the matter source gravitational wave corrections can dominate over the direct interaction between the EMRI and the matter. These works, however, obtain the direct matter interaction from dynamical-friction prescriptions~\cite{Rahman:2025mip} or asymptotic matter fluxes~\cite{Keijzer:2026vul,Li:2026foy}. An independent calculation of these contributions using our first-principles forcing functions, Eq.~\eqref{eq:g01_final_1PA}, would provide an important check of how these and other works connect asymptotic matter and gravitational-wave fluxes to the secondary's dynamics through assumed balance laws.

We believe our results in this section clarify the current state of the field in several key ways. We derive from first principles how matter torques acting on a secondary body arise in curved spacetime, and identify the radiative contribution from the perturbed matter field, correcting the acausal construction used in some previous works. We further obtain all of the purely gravitational contributions that enter at $\mathcal{O}(\epsilon^2\lambda)$, which are similar to those calculated in Refs.~\cite{Rahman:2025mip,Keijzer:2026vul,Li:2026foy}, but in our case arise from the conservation of the point-particle stress-energy tensor on the appropriately regularised metric. Most importantly, our formulae are constructed entirely from much more easily calculable quantities, so that for environments that are stationary and axisymmetric at leading order, the full 1PA environmental forcing of the energy and angular momentum only requires solving for the perturbed matter response, $T^{\mu\nu}_{(\env,1,1)}$, and the leading-order point-particle, $h^{(\rm p,1,0)}_{\mu\nu}$, and environmental, $h^{(\env,0,1)}_{\mu\nu}$, metric perturbations.

\section{Conclusion}

In this work we have developed a relativistic form of Newton's Third Law, which we hope will provide a straightforward means to construct forcing functions for EMRI orbital evolution in relativistic environments. Motivated by the Newtonian Third Law, our goal was simple: to rewrite the force exerted by the environment on the EMRI in a form resembling the integrated work or torque the EMRI applies to the environment, thereby avoiding the need to explicitly construct the environment's dynamical metric perturbation. We achieved this by constructing quasi-conserved orbital charges and exploiting the symmetries of Green's functions in the presence of Killing symmetries. 

We expect practical applications of the ``instantaneous'' Third Law derived in Eq.~\eqref{eq:BalanceLaw}, built from the time-localised advanced point-particle metric perturbation, to reveal rich features of the causal interaction between dynamical matter fields and EMRIs during inspiral. Beyond this, we have performed a long-term orbit average of the instantaneous results in Eq.~\eqref{eq:averaged 3rd law v1}, revealing the relativistic Third Law in the averaged sense. We find that, on average, the work or torque exerted on the EMRI by the environment is equal and opposite to the work or torque exerted on the environment by the EMRI, up to the mixed fluxes of energy and angular momentum generated jointly by the EMRI and the matter. These flux corrections can be written explicitly in terms of gravitational-wave amplitudes at infinity and on the primary's horizon, obtained by integrating the secondary's and environment's stress-energy tensors against the radiative Green's function of Kerr. For environments that are dynamical at leading order, this flux term already enters at $\mathcal{O}(\epsilon\lambda)$. Depending on the size of $\lambda$, it can therefore exceed the leading-order vacuum gravitational-wave flux, which enters at $\mathcal{O}(\epsilon^2)$. Only in the weak-field limit, where the radiative field vanishes and these fluxes become subdominant, does one recover the na\"ive, acausal Newtonian result in the averaged sense.

Moving to the direct applications, we have provided a practical formulation of our Third Law in two distinct settings, which we see as its two primary use-cases. The first is an averaged form at leading order, Eq.~\eqref{eq:averagedequationsenvdecomp}, suitable for EMRI evolutions in highly dynamical environments with rich temporal structure. This construction requires only the leading-order point-particle metric perturbation and the matter field evolved on $g^{\rm Kerr}_{\alpha\beta}+h^{(\rm p)}_{\alpha\beta}$. The second is a form at first post-adiabatic order, Eq.~\eqref{eq:g01_final_1PA}, for environments whose leading-order configurations are stationary and axisymmetric. This construction provides a forcing function applicable to most self-consistent matter--EMRI calculations currently of interest in the relativistic literature. It requires only the matter response and the leading-order point-particle and environment metric perturbations, given by $T^{\mu\nu}_{(\env,1,1)}$, $h_{\mu\nu}^{(\rm p,1,0)}$, and $h_{\mu\nu}^{(\env,0,1)}$ respectively, avoiding the calculation of any gravitational perturbations at $\mathcal{O}(\epsilon\lambda)$. We expect our $1$PA results to be relatively straightforward to generalise to stationary, axisymmetric environments that accrete onto the primary, such as viscous discs~\cite{Potter_2021}, in which the background matter flow $T^{(\env,0,1)}_{\alpha\beta}$ is stationary and axisymmetric but the resulting metric perturbation $h^{(\env,0,1)}_{\alpha\beta}$ is only axisymmetric. Extensions to matter profiles without these symmetries, however, remain unclear, both in the application of the conservation of the point-particle stress-energy tensor and in the treatment of the leading-order resonances that would arise as in Sec.~\ref{sec:BalanceTypeD}. We hope these results will provide a concrete and considerably simpler means of constructing precision LISA waveform models for EMRIs in environments than the approaches currently taken in the literature.

Our formulation of the relativistic Third Law relies on the existence of Killing vectors, meaning (for a Kerr background) it describes the evolution of energy and angular momentum. Since the Carter constant arises from a Killing tensor rather than a Killing vector, our results do not yet extend to its evolution. As a consequence, even though our results are valid for generic orbital configurations, they do not provide a complete prescription for the evolution of inclined orbits. Extending them to the evolution of the Carter constant is an important direction for future work, perhaps utilising tools from Refs.~\cite{Isoyama:2018sib,Grant:2024ivt,Lewis:2025ydo}. On the other hand, the same reliance on Killing symmetries means that our construction carries over straightforwardly to the post-Minkowskian setting at leading order, where the background is flat and the Killing vectors correspond to the translations, rotations, and boosts of the Poincar\'e group. Specialising to the spatial translations, whose conserved charges are the linear momenta, gives balance laws for the change in linear momentum, i.e.\ the Cartesian force. This yields a leading-order relativistic Third Law for the full set of spatial forces in spacetimes that are flat at leading order. As such our results are directly applicable to the early inspiral of gas-embedded massive black hole binaries, which, alongside EMRIs, are key targets of the upcoming LISA mission. We provide an application of the Third Law at leading order in this context in Appendix~\ref{app:PM_third_law}. 

Beyond simplifying the calculation we expect our results to clarify some long-standing ambiguities in the community regarding the proper treatment of EMRI-environment evolution. We do this by connecting the force on the EMRI to more readily available quantities which resemble, but are not equivalent to, the work and torque the EMRI applies to the environment. One important direction we do not tackle here is how the integrated matter forces can be formally connected to asymptotic matter fluxes, in the subset of environment models that allow it. Such connections have recently been made in the relativistic literature for spiral-arm calculations of EMRIs in thin discs~\cite{Dyson:2026ddd,Dittmann:2026rtj}, but a more general, formal connection remains an interesting open problem. Finally, we believe an important future step, and check of the results presented here, will be to calculate, from first principles, the direct forces on the EMRI from the environment as predicted by the local regular field, $h^{({\rm R},1,1)}_{\alpha\beta}$, in self-force theory. This will require careful treatment of the singularity structure of the secondary and, importantly, a first-principles matched asymptotic analysis of the coupled EMRI-environment system.

\acknowledgments

C.D. is grateful to Richard Brito and Robrecht Keij\-zer for insightful discussions throughout the course of this work.
C.D. is funded by the European Union (ERC Starting Grant, Unleash-TDEs, project number 101163093). Views and opinions expressed are however those of the author(s) only and do not necessarily reflect those of the European Union or the European Research Council. Neither the European Union nor the granting authority can be held responsible for them.
The Center of Gravity is a Center of Excellence funded by the Danish National Research Foundation under grant No.\ 184. 
C.D. and M.v.d.M. \ acknowledge support by VILLUM Foundation (grant no.\ VIL37766 and no.\ VIL53101)
and the DNRF Chair program (grant no.\ DNRF162) by the Danish National Research
Foundation. 
M.v.d.M. is  supported by the European Union’s Horizon ERC Synergy Grant ``Making Sense of the Unexpected in the Gravitational-Wave Sky'' (GWSky--101167314).
D.J.D.\ acknowledges support from the STScI Director's Discretionary Research Fund.
L.Z. is supported by the European Union’s Horizon 2024 research and innovation program under the Marie Sklodowska-Curie grant agreement No. 101208914. 
A.P. acknowledges the support of the ERC Consolidator/UKRI Frontier Research Grant GWModels (as selected by the ERC and funded by UKRI under grant number EP/Y008251/1). The authors made use of Claude (Anthropic) to assist in converting a hand-drawn schematic to TikZ code for Fig.~\ref{fig:Worldline}, and to derive the proof that the structure of the first-order puncture implies it is polyharmonic, as stated in footnote~\ref{fn:polyharmonic}, which the authors subsequently checked.

\appendix

\section{Comparable-mass binaries in the weak field}\label{app:PM_third_law}
Our construction in  Secs.~\ref{sec:Formalism} and \ref{sec:BalanceFirst} relies on two key properties: the existence of some Killing symmetries on the background spacetime, and the construction of linearised metric perturbations from Green's functions satisfying Eqs.~\eqref{eq:relativistic identity 1} and~\eqref{eq:relativistic identity 2}. As neither of these properties are unique to the problem of studying EMRIs, in this appendix we extend the relativistic Third Law to comparable-mass binaries embedded in environments at leading order in the gravitational constant ($G$). We expect this construction to be particularly relevant for the study of massive black hole binaries embedded in discs during their early stages in the LISA band. Throughout this work we have set units such that $G=1$; hence, as with the previous perturbative parameters, we treat $G$ as an order-one counting parameter. 

We begin by considering the form of \eqref{eq:full_ein} now for two compact objects of mass $m_A$, $A\in\{1,2\}$, whose non-linear worldlines are given by $z^{\alpha}_{A}(\tau_A)$ with corresponding four-velocities $u^{\alpha}_{A}$, embedded in an environment on an approximately Minkowski spacetime. This gives the sourced Einstein equations in the form
\begin{equation}
    G_{\mu\nu}[\bm{g}] = 8 \pi G \left( T^{({\rm p}_1)}_{\mu\nu} + T^{({\rm p}_2)}_{\mu\nu} + \lambda T^{(\env)}_{\mu\nu}\right),
\end{equation}
where we treat the worldlines $z^\alpha_A(\tau_A)$ appearing in $T^{({\rm p}_1)}_{\mu\nu}$ and $T^{({\rm p}_2)}_{\mu\nu}$ as prescribed: they may be bound or scattering and are retained to all orders in velocity, with their acceleration due to the companion arising at $\mathcal{O}(G)$. Our interest is only in the additional force due to the environment, which enters at $\mathcal{O}(G\lambda)$. At leading order in the post-Minkowskian expansion, we can then take the metric to be given by
\begin{equation}
    \bm{g}_{\mu\nu} = \eta_{\mu\nu} + G \left( h^{(\mathrm{p}_1)}_{\mu\nu}+ h^{(\mathrm{p}_2)}_{\mu\nu}+\lambda h^{(\env)}_{\mu\nu}\right)+\cdots.
\end{equation}
Motivated by the fact that, once in-band, massive black hole binaries will generally be in the nonlinear matter-response regime, we keep our analysis to leading order in this section. 

Following the previous approach we construct our metric perturbations through integration against the Minkowski Green's function:
\begin{equation}
    h^{(\rm \mathrm{p}_A/\env)}_{\alpha\beta} = \int G^{\rm Ret}_{\alpha\beta\alpha'\beta'}(x,x')T^{\alpha'\beta'}_{(\mathrm{p}_A/\env)}(x')d^4x'.
\end{equation}
Writing the fields in terms of global inertial coordinates, $x^{\alpha} = (t,\bm{x})$, we can express the Lorenz-gauge Green's function  explicitly as
\begin{align}
    G^{\rm Ret/Adv}_{\alpha\beta\alpha'\beta'}(x,x') = 4\bigg(\eta_{\alpha(\alpha'}&\eta_{\beta')\beta}-\frac{1}{2}\eta_{\alpha\beta}\eta_{\alpha'\beta'}\bigg)\label{eq:Minkowski Green's function}\\ \notag
    &\times \frac{\delta 
    (t-t'\mp|\bm{x}-\bm{x}'|) }{|\bm{x}-\bm{x}'|},
\end{align}
which can be shown to satisfy the identities in Eqs.~\eqref{eq:relativistic identity 1} and \eqref{eq:relativistic identity 2}. 

Next, we again take hypersurfaces $\Sigma_t$, now parameterised by the inertial time coordinate. Based on these coordinates $(t,\bm{x})$, we consider the Killing vectors associated with the translation symmetries of the system, denoting them by $\xi^{\nu}_{(\mu)}=\delta^\nu_\mu$. Our quasi-conserved charges are then
\begin{equation}
    P_{\mu}(\Sigma_t):= \int_{\Sigma_t}T^{\ \alpha}_{\mu} \,d\Sigma_{\alpha},
\end{equation}
which, at leading order for each body, simply gives its linear momentum, $P_{\mu}^{(A)}= m_Au^A_{\mu}$. At leading order in the environmental force, following the same steps as in Sec.~\ref{sec:BalanceFirst}, we then arrive at an explicit form of the direct force on the objects given by
\begin{align}
    F^{(A)}_{\mu}:= \frac{d P_{\mu}^{(A)}}{dt} &= \frac{1}{2}\int_{\Sigma_t}T^{\alpha\beta}_{(\mathrm{p}_A)}\partial_{\mu}h^{(\env)}_{\alpha\beta}d\Sigma\\ \notag
    &= \frac{m_A}{2u^t_A}u^{\alpha}_{A}u^{\beta}_{A}\partial_{\mu}h^{(\env)}_{\alpha\beta}\bigg|_{z_{A}(t)}.
\end{align}

With this construction the application of the Third Law remains unchanged from Sec.~\ref{sec:instantaneous}, and we can build the instantaneous metric perturbation whose source is given by
\begin{equation}T^{\alpha\beta}_{(\mathrm{p}_A|t^*)}:= \frac{m_A}{u^t_A}u^{\alpha}_Au^{\beta}_A\delta^4(x^\mu-z^{\mu}_A(t^*)).
\end{equation}
Integrating against the advanced Green's function~\eqref{eq:Minkowski Green's function}, one then finds an explicit form of the instantaneous metric perturbation~\eqref{eq:localised hAdv} for each particle, given by 
\begin{align}
    h^{(\mathrm{p}_A,\mathrm{Adv}|t^*)}_{\mu\nu}
    = \frac{4 m_A}{u^t_{A}}\bigg(&u^{A}_{\mu}u^{A}_{\nu}+ \frac{1}{2}\eta_{\mu\nu}\bigg)\notag\\
    &\times\frac{\delta(t-t^* + |\bm{x}-\bm{z}_A(t^*) |)}{|\bm{x}-\bm{z}_A(t^*)| },
\end{align}
where $u^\alpha_A$ is evaluated at the source point $z_A(t^*)$. Since it is independent of the field point $x$, the derivative $\partial_\mu$ in Eq.~\eqref{eq:Instnat_3Law_PM} acts only on the delta-function factor. This gives rise to the relativistic Third Law on flat space, in analogy with Eq.~\eqref{eq:BalanceLaw}:
\begin{align}\label{eq:Instnat_3Law_PM}
    \frac{dP^{(A)}_{\mu}}{dt^*}
    &= -\frac{1}{2}\int d^4x\; T^{\alpha\beta}_{(\rm env)}(x)\,\partial_{\mu}h^{({\rm p}_A,{\rm Adv}|t^*)}_{\alpha\beta}(x)\\ \notag
    &= -\frac{m_A}{u^t_A}\int d^4x\; \rho^{\rm act}_A\partial_{\mu}\!\left[\frac{\delta\big(t-t^*+|\bm{x}-\bm{z}_A(t^*)|\big)}{|\bm{x}-\bm{z}_A(t^*)|}\right],
\end{align}
where we have defined
\begin{equation}
    \rho^{\rm act}_A:=T^{\alpha\beta}_{(\env)}\left(2u^A_{\alpha}u^A_{\beta}+\eta_{\alpha\beta}\right),
\end{equation}
the active gravitational mass density of the environment as measured by an observer comoving with object $A$, which reduces to $\rho+3p$ for a perfect fluid at rest relative to $A$. Integrating by parts and assuming the matter field decays fast enough not to contribute boundary terms, one can remove the delta function by performing the time integration to give
\begin{equation}\label{eq:Third Law for comparable masses}
    \frac{d P^{(A)}_{\mu}}{dt^*} = \frac{m_A}{u^t_A}\int d^3x\,\frac{\partial_{\mu}\rho^{\rm act}_{A}\big(t^*-|\bm{x}-\bm{z}_A(t^*)|,\bm{x}\big)}{|\bm{x}-\bm{z}_A(t^*)|}.
\end{equation}
Here $\partial_{\mu}\rho^{\rm act}_A$ is evaluated at the retarded time $t_{\rm ret} = t^*-|\bm{x}-\bm{z}_A(t^*)|.$ 
Equation~\eqref{eq:Third Law for comparable masses} is an explicit expression for the force on the compact object A written as an integral over the energy content of the matter field.

\bibliography{emri_disk_torques}


\end{document}